\documentclass[iop,twocolumn]{aastex701}
\usepackage{graphicx,multirow,hyperref,url,color,amsmath}
\usepackage{makecell}
\usepackage{tabularx,booktabs,threeparttable}
\usepackage[normalem]{ulem}
\usepackage{CJKutf8}
\usepackage{lineno}

\hypersetup{
  colorlinks,
  linkcolor={blue!88!black!80},
  citecolor={blue!88!black!80},
  urlcolor={blue!88!black!80}}

\newcommand{\msun}{{M_\odot}}
\newcommand{\mbh}{M_{\rm BH}}
\newcommand{\ergs}{{\rm erg\,s^{-1}}}

\newcommand{\asep}{a_{\rm sep}}

\newcommand\yps{\bgroup\markoverwith{\textcolor[rgb]{1.0, 0.0, 1.0}{\rule[0.5ex]{8pt}{1.5pt}}}\ULon}

\begin{document}
\begin{CJK*}{UTF8}{gbsn} 

\shorttitle{Evolution and SED of bMBH Accretion}
\title{The Distinctive Evolution and Spectral Energy Distribution of Binary Massive Black Hole Accretion}

\author{Yue Xu (许悦)}
\affiliation{Shanghai Astronomical Observatory, Chinese Academy of Sciences, 80 Nandan Road, Shanghai 200030, People's Republic of China}
\affiliation{School of Astronomy and Space Sciences, University of Chinese Academy of Sciences, 19A Yuquan Road, Beijing 100049, People's Republic of China}
\email[]{xxx@}

\author[0000-0001-9969-2091]{Fu-Guo Xie (谢富国)}
\altaffiliation{fgxie@shao.ac.cn}
\affiliation{Shanghai Astronomical Observatory, Chinese Academy of Sciences, 80 Nandan Road, Shanghai 200030, People's Republic of China}
\email[]{fgxie@shao.ac.cn}

\author[0000-0002-7329-9344]{Ya-Ping Li (李亚平)}
\affiliation{Shanghai Astronomical Observatory, Chinese Academy of Sciences, 80 Nandan Road, Shanghai 200030, People's Republic of China}
\email[]{xxx@}

\author[0000-0003-3728-8231]{Yinhao Wu (吴寅昊)}
\altaffiliation{EACOA Fellow}
\affiliation{Shanghai Astronomical Observatory, Chinese Academy of Sciences, 80 Nandan Road, Shanghai 200030, People's Republic of China}
\email[]{xxx@}

\author[0000-0001-6947-5846]{Luis C. Ho (何子山)}
\affiliation{Kavli Institute for Astronomy and Astrophysics, Peking University, Beijing 100871, People's Republic of China}
\affiliation{Department of Astronomy,
School of Physics, Peking University, Beijing 100871, People's Republic of China}
\email[]{xxx@}

\date{Accepted XXX. Received YYY; in original form ZZZ}

\begin{abstract}
Binary (super-)massive black holes (BHs) are expected to reside in the center of some galaxies. In this work, we re-visit accretion onto binary massive BHs, incorporating recent advances in both accretion theory and the mass transfer rate between the two massive BHs. We focus on relatively bright systems with an Eddington ratio of 0.1 for a binary with total BH mass $10^8\,M_\odot$, but consider a wide range of mass ratios $10^{-4} \le q \le 0.5$. The binary system consists of two mini-disks surrounding two individual BHs and a circumbinary disk surrounding the mass center of binary BHs. Depending on the mass ratio, the two mini-disks can be hot accretion flows, standard thin (cold) disks, or Slim disks. The radiative contributions from all three disks, each potentially in different accretion modes, are taken into account self-consistently. The spectral energy distributions of the binary BH system show universal ``notch'' features from the near-infrared to ultraviolet bands, caused by the gap or cavity in the accretion disk, consistent with previous studies. Binary with different mass ratios exhibit distinct spectral energy distribution properties, offering opportunities for testing (identifying candidates) with future broad band (infrared up to X-rays) observations. We also investigate the evolution of these binary systems, and find that, for systems with initial mass ratios $q \lesssim \text{a few} \times 10^{-3}$, the mass ratio evolves toward an equilibrium value $q \sim 10^{-3}$. For binary BH systems with a larger initial mass ratio, their mass ratio instead evolves toward unity.
\end{abstract}




\section{Introduction}\label{sec:intro}

It is now widely accepted that supermassive black holes (SMBHs)—defined here as those with masses $\mbh\ga10^6\msun$—ubiquitously exist in the centers of nearly all individual galaxies \citep{Kormendy2004, kormendy2013}. Under the hierarchical framework of galaxy formation and evolution within modern cosmology, there could be more than one (super-)massive black hole (BH) in the center of each galaxy (e.g., \citealt{Graham2004, Merritt2007, Volonteri2022}; see \citealt{silk2012} for a review). 
Galaxies undergo numerous merger events. Following a galaxy-galaxy merger, the SMBHs from the progenitor galaxies will settle into the center of the newly formed gravitational potential, resulting in a phase where two SMBHs coexist in the post-merger system. Once the separation between the two SMBHs decreases to about 1 parsec via dynamical friction by gas/stars and stellar ``loss-cone'' scatterings, they may form a gravitationally bound, interacting binary SMBH system (e.g., \citealt{2007Mayer, Fu2018, Kelley2019, Chen2020}) that subsequently inspirals inward prior to their eventual coalescence.

Besides the aforementioned galaxy-galaxy merger pathway, another route to form binary massive BHs arises from dynamical processes within individual galaxies. Intermediate-mass black holes (IMBHs; defined as those with $\mbh \sim 10^{2-6}\msun$, see \citealt{Greene2020} for a review) may form via collisions and subsequent mergers of stars and stellar-mass black holes in dense clusters (e.g., \citealt{2004pz, 2004Gurkan, 2021dicarlo, 2021rizzuto}; see the review by \citealt{Greene2020}), or from the prompt evolution of supermassive star, which is formed by a rapid collapse of chemically pristine primordial gas in so-called atomic-cooling halos (see discussions in \citealt{Inayoshi2020}). Super-Eddington accretion of massive stellar BHs can also form IMBHs. LIGO-Virgo observations of gravitational waves from mergers of stellar BHs also provide direct evidence on the formation of $\sim 100-300 \msun$ ``Lite'' IMBHs \citep{Abbott2020, RR2025}. Dynamical friction can cause the orbital decay of globular clusters toward the nucleus of the host galaxy, where the IMBH may eventually be captured by the SMBH to form an IMBH-SMBH binary system (e.g., \citealt{2001ebis}). This process is expected to be more efficient if the SMBH is surrounded by an accretion disk (i.e., the nucleus hosts a active galactic nucleus; AGN), and the IMBH interacts with this disk.
Close binary massive BHs, both binary SMBHs or IMBH-SMBH systems, are of particular interest because they emit gravitational waves during their inspiral and merger, which are detectable by pulsar timing arrays (e.g., \citealt{2013sesana}) and space-based interferometers such as LISA \citep{2023AS}, TianQin \citep{Luo2016} and Taiji \citep{Ruan2020}.

Observationally, very long baseline interferometry (VLBI) in the radio band provides a powerful tool for detecting or confirming binary SMBHs with separations at $0.1$ parsec level (e.g., \citealt{2006Rodriguez, DeRosa2019, 2023jiang}). For binary SMBHs with smaller separations, current facilities cannot resolve them spatially. Their presence can be investigated spectroscopically. Several classical methods have been developed for this purpose. One approach is to search for velocity shifts in the broad emission lines, which vary as the binary components orbit each other (e.g., \citealt{Gaskell1996, Eracleous2012, Ji2023}). Another method involves looking for bimodality in the narrow-line profiles (e.g., \citealt{Shen2011, Comerford2013, Ji2023}). The other technique is to identify periodic variations in the light curve, which arise from the relativistic Doppler effect as the binary BH orbit and periodically amplify or diminish the observed luminosity \citep{Lodato2009}. This method has been used to identify binary SMBH candidates through long-term photometric monitoring (e.g., \citealt{Charisi2018, D'Orazio2015, Zheng2016, Ji2021}). Finally, it has also been suggested that reverberation mapping can be used to identify binary broad line regions of binary SMBHs, and thus provides a method to find binary SMBH candidates \citep{Songsheng2020}.

Recently, several new observational methods for binary BHs have been proposed. It has been found that when the binary orbit is viewed nearly edge-on, the gravitational lensing effect of the foreground black hole amplifies the radiation from the background mini-disk (cf. Sec.~\ref{sec:model}), resulting in a periodic, sharp increase in flux (e.g., \citealt{Krauth2024}). 
\citet{Fu2025} proposes a new method for identifying low-mass-ratio binary SMBHs. The key principle is that the low-density cavity within the binary's accretion flow produces a distinctive break in the inter-band time lag versus wavelength relation, $\tau(\lambda)$. Specifically, the relation appears flat at short wavelengths due to emission from the truncated mini-disks, transitioning to a $\lambda^{4/3}$ power-law at long wavelengths originating from the circumbinary disk. This characteristic signature fundamentally distinguishes accretion onto binary SMBHs from accretion onto single SMBH. Furthermore, they also applied this technique to a candidate binary SMBH source PG 1302-102.

\begin{figure*}
\centering
  \includegraphics[width=0.45\textwidth]{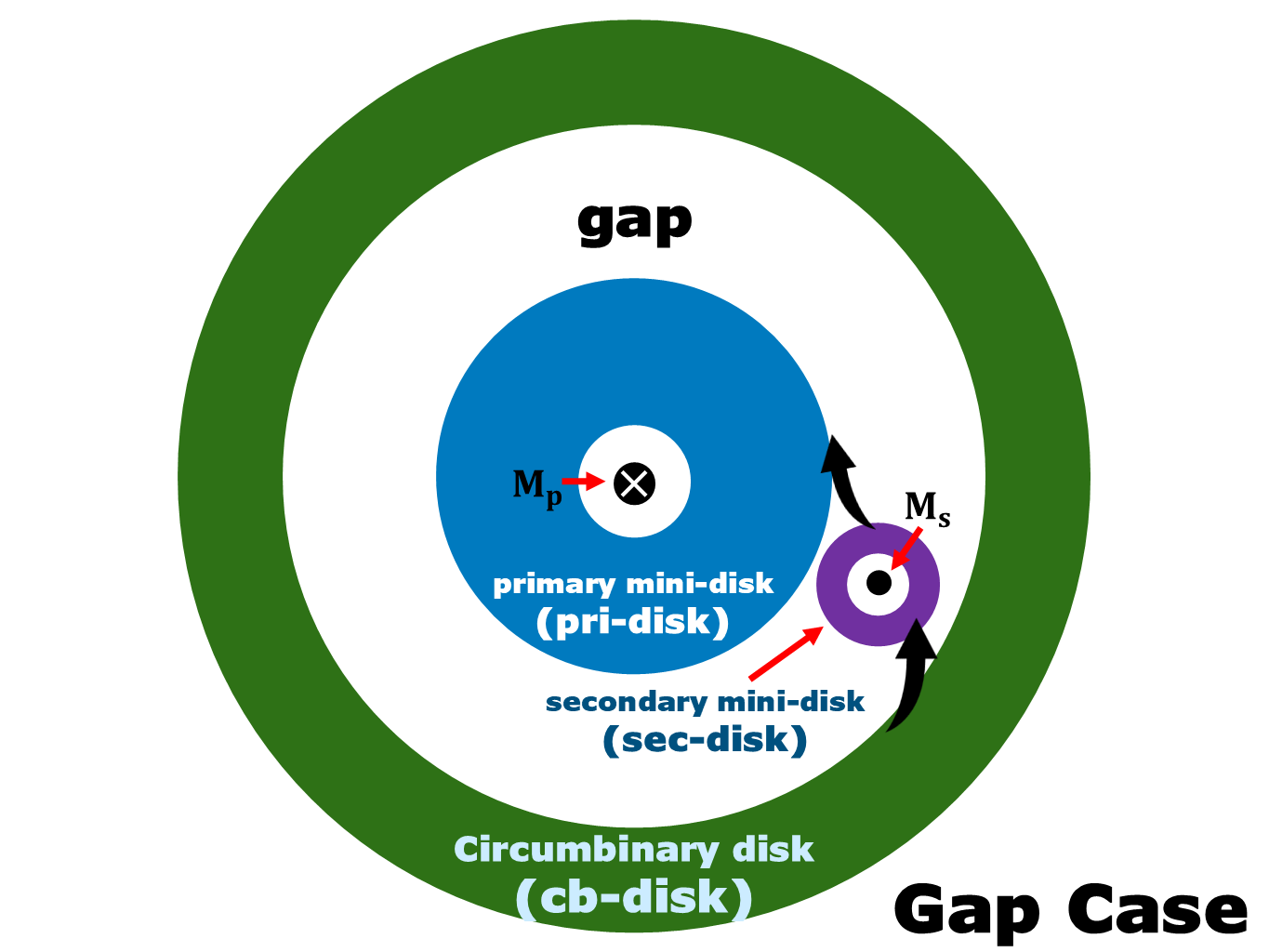}
  \hspace{0.8cm}\includegraphics[width=0.45\textwidth]{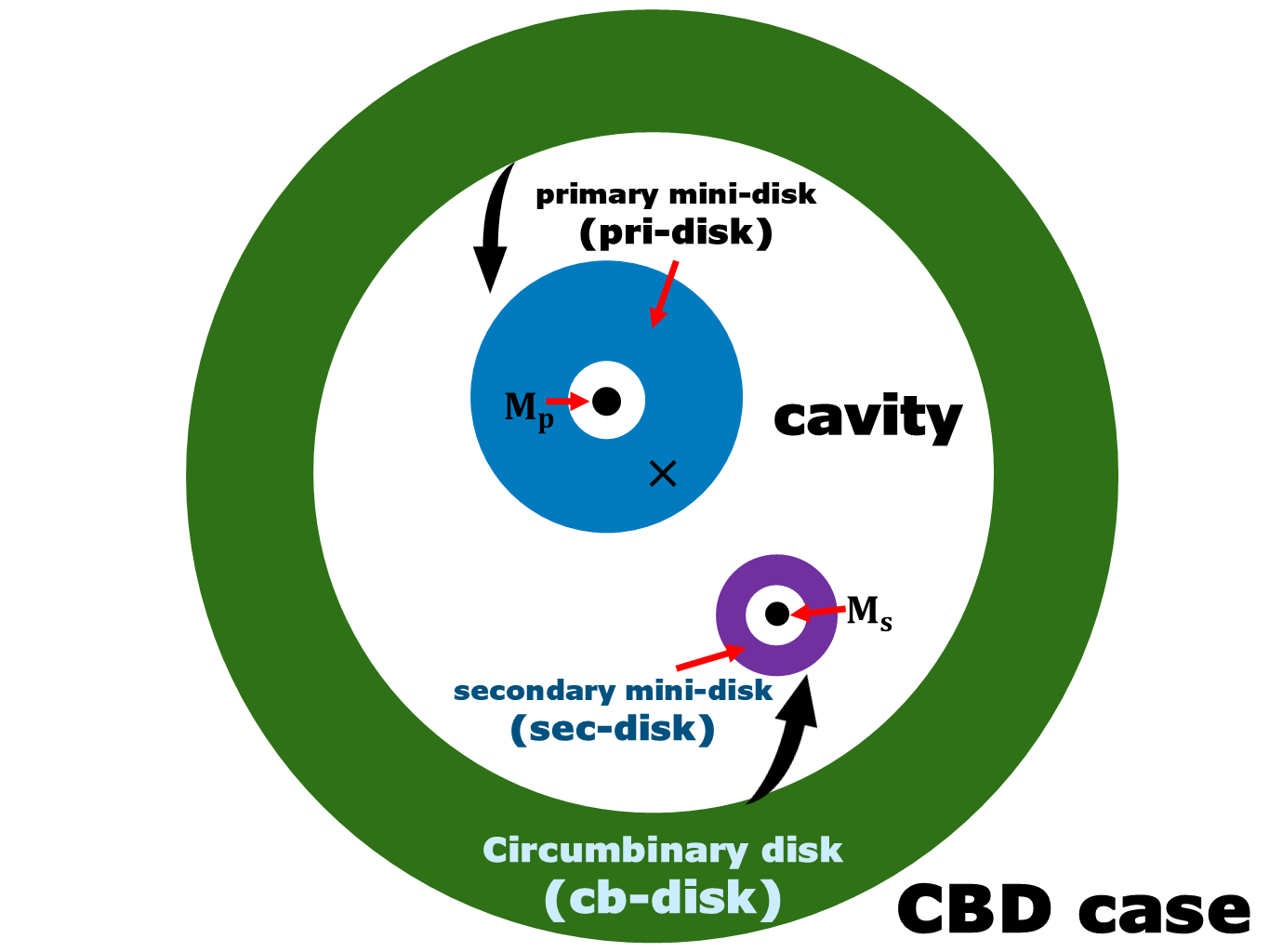} 
  \caption{A schematic plot of accretion onto binary massive BHs. {\it Left panel}: the ``gap case'', where a low-mass-ratio secondary BH clears a narrow annular gap in the disk. {\it Right panel}: the ``circumbinary disk (CBD) case'', where the binary BHs opens a wider central cavity. In both panels, the two BHs (shown by black ``$\bullet$''s) are separated by a distance $a_{\rm sep}$, and orbit their mass center (shown as ``$\times$'' mark, for the gap case, the mass center is very close to the primary) on circular (zero-eccentricity) trajectories. The system comprises three main accretion structures: an outer circumbinary disk (green ring，``\texttt{cb-disk}''), and two inner mini-disks that individually surround the primary (blue ring, ``\texttt{pri-disk}'') and secondary (purple ring, ``\texttt{sec-disk}'') massive BHs. These components are connected by accretion streams, indicated by the arrows.}
  \label{fig:bbh_schem}
\end{figure*}

In this work, we focus on the continuum emission from  binary massive BHs with different mass ratios in which the primary is an SMBH. We consider galaxies in an active phase, i.e., those hosting an AGN with an accretion disk surrounding the primary BH (for a comprehensive review of AGNs, see \citealt{Padovani2017}). We assume that the secondary BH settles within the accretion disk of the primary SMBH, such that each BH is surrounded by its own accretion disk.
Many works have applied the standard thin accretion disk \citep[][i.e., Shakura-Sunyaev disk, hereafter SSD]{Shakura1973} to model the accretion physics in binary massive BH systems (e.g., \citealt{Artymowicz1994, Artymowicz1996, Hayasaki2008, Noble2012}). 
However, it is widely believed that there is a broad and deep cavity around the binary orbit in the binary SMBH case \citep[see review by][]{Lai2023}.
For a wide range of mass ratio among the IMBH-SMBH system, a gap will form in the AGN disk (cf. Section~\ref{sec:model}). It can cause a depression of its spectrum in the near-infrared up to optical-ultraviolet band, compared to that of the standard AGN disk (e.g., \citealt{Sesana2012, Tanaka2012, Tanaka2013, Farris2015, Yan2015, Zheng2016}), which typically is optically-thick but geometrically-thin \citep[][]{Shakura1973}.

This work is organized as follows: In Section~\ref{sec:model}, we describe the binary massive BH accretion model we adopted. In Section~\ref{sec:rad}, we model the spectral energy distributions (SEDs) of accretion onto binary massive BHs, where impacts of BH separation and mass ratio are considered. In Section ~\ref{sec:evol}, we consider the distinctive mass transfer rate among two massive BHs and investigate their long-term evolution in secondary-to-primary BH mass ratio $q$. Finally, we discuss the observational implications and limitations in Section~\ref{sec:discussion}, and provide a brief conclusion in Section~\ref{sec:conclusion}.

\section{Model of Accretion onto Binary Massive BHs}\label{sec:model}

\subsection{Basic Assumptions}\label{sec:assumptions}

In this work, we assume the total BH mass of the system is $M_{\rm bin} = M_{\rm p} + M_{\rm s} = (1+q)M_{\rm p}$, where the masses of the primary and the secondary BHs are denoted as $M_{\rm p}$ and $M_{\rm s}$, respectively. $q=M_{\rm s}/M_{\rm p}$ (note that $q\leq 1$) defines the mass ratio. 
We consider the regime with $10^{-4}\lesssim q\lesssim 1$, the case with $q\ll 1$ can be regarded as a binary with an extreme (or intermediate) mass ratio secondary of the IMBH-SMBH system, and $0.01\lesssim q\lesssim 1$ is the classical binary SMBH system. 
The coordinate center locates at the mass center of two BHs. 
For convenience, we set the BH spin ${a}^{\star}=0.5$ and the orbital eccentricity of the binary $e=0$.
The angular velocity of the binary system is then derived from Kepler's law, $\Omega_{\rm bbh}^{2} = \frac{GM_{\rm bin}}{a_{\rm sep}^3}$, and the orbital period of the binary BHs is $P_{\rm bin} = 2\pi/\Omega_{\rm bbh}$. In this setup, the separation of binary BHs has a fixed value of $a_{\rm sep}$. The primary and the secondary BHs locate at, respectively, $R_{\rm pri}=\frac{qa_{\rm sep}}{1+q}$ and $R_{\rm sec}=\frac{a_{\rm sep}}{1+q}$. We further define $R_{\rm g} = GM_{\rm bin}/c^2$ as the ``effective'' gravitational radius of the binary.

As we will show below, the accretion onto the binary system can be described into three disks, i.e. a circumbinary disk, and two mini-disks around each BH. The three disks are assumed to be mutually aligned and share the same prograde direction as the binary orbit.
The viscosity parameter of all the three accretion disks is set to $\alpha_{\rm vis}=0.3$. We adopt a simple model to approximate the continuous radiation of the binary BH accretion system as a combination of the radiation from the outer circumbinary disk and the two mini-disks around the binary BHs. Assuming an accretion rate of $0.1\dot{M}_{\rm Edd,bin}$ into the binary, the circumbinary disk is described by an SSD model, whose radiation is approximated by polychromatic blackbody radiation (\citealt{Shakura1973, Novikov1973}). Here, the Eddington luminosity and accretion rate of a $\mbh$-mass BH are, $L_{\rm Edd}\approx 1.3\times10^{46}\,\ergs\, (\mbh/10^8\msun)$ and $\dot{M}_{\rm Edd} = L_{\rm Edd}/(0.1 c^2)$, respectively. A dimensionless accretion rate (in Eddington unit) can be defined as $\dot{m}=\dot{M}/\dot{M}_{\rm Edd}$. 
We further assume a line-of-sight inclination angle of $i=30^{\circ}$ to calculate the radiation from the system.

In the following, we will explore the structure and radiative properties of accretion onto binary massive BHs (both binary SMBH and IMBH-SMBH).

\subsection{Model Configuration: Gap Case versus CBD Case}

Consensus has been reached recently (especially in the protoplanetary community) that, depending on the mass ratio of the binary BHs and the disk properties, there will be two distinctive configurations for the binary system, as shown in Figure \ref{fig:bbh_schem}. The left panel illustrates the extreme mass-ratio system (e.g., $q \ll 1$ and $q \gtrsim 10^{-6}$), referred to as the ``gap case'', which is analogous to a planet embedded in a protoplanetary disk. The right panel shows the case with  $q \lesssim 1 $, corresponding to the ``CBD case''. Below we will provide more details on each configuration.

\subsubsection{Gap case at $q\ll1$}

For the extreme mass ratio case of IMBH-SMBH system, the gravitational interaction between the binary and disk is similar to the planet-disk system.
In the classical theory of planet-disk interactions, the ability of a planet to open a significant gap in a protoplanetary disk is determined by thermal and viscosity criteria \citep[e.g.,][]{Lin1993,Crida2006}. However, recent numerical simulations suggest that the thermal criterion may not be necessary and the condition for the opening of the gap is simply \citep{Duffell2013, Zhu2013, Kanagawa2015, Paardekooper2023},
\begin{equation}
    K\equiv \alpha_{\rm vis}^{-1}h^{-5}q^2 \gtrsim 25, \label{eq:qcrit}
\end{equation}
where $h \equiv H/R$ defines the height-to-radius aspect ratio of the disk. We emphasize that, unlike protoplanetary disks whose $h \sim 0.05-0.1$, the (cold) accretion disks in AGNs typically have $h\sim 10^{-3}-0.01$. Consequently, a lite secondary BH, with a very small mass ratio $q\sim 10^{-5} - 10^{-4}$ (i.e., a typical IMBH secondary, for a $10^8\msun$ primary), still can efficiently open a gap in its host AGN disk. Hereafter we define this as the ``gap case''. The accretion structure in this case consists of three disks as shown in the left panel of Fig.\ref{fig:bbh_schem}, i.e. an outer disk (we term it as circumbinary disk, ``\texttt{cb-disk}'' hereafter, shown as a green ring), and two mini disks around， the primary (``\texttt{pri-disk}'' hereafter, blue ring) and the secondary (``\texttt{sec-disk}'' hereafter, purple ring) BHs, respectively. Accretion streams, shown by arrows in the figure, indicate the connections between the three accretion disks. 

The outer truncation radii of the secondary mini-disks is set by the Hill spheres of the respective BHs $R_{\rm out} \approx 0.4\,R_{\rm Roche}$ (\citealt{Martin2011}), where an approximate expression for the Roche radius is given by \citep{Eggleton1983},
 \begin{equation}
    R_{\mathrm{Roche}} = a_{\mathrm{sep}} \cdot \frac{0.49 \, q^{2/3}}{0.6 \, q^{2/3} + \ln\left( 1 + q^{1/3} \right)}.
    \label{eq:roche}
\end{equation}
A key parameter to describe the substructures of the disk is the gap width, denoted as $W_{\rm gap}$. In previous studies, these quantities differed slightly in definition, although they were somewhat similar in size. The gap is maintained by balancing the viscous stress with the gravity torque driven by the secondary BH. We determine the gap width using the empirical scaling law derived from numerical simulations by \citet{Kanagawa2016}:
\begin{equation}
    \frac{W_{\rm gap}}{a_{\rm sep}} = 0.41 K'^{1/4},
    \label{eq:gapwidth}
\end{equation}
where the dimensionless parameter is defined as $K' = \alpha_{\rm vis}^{-1} h^{-3} q^2$. Clearly, a lower aspect ratio (cooler disk) and/or a larger mass ratio $q$ (more massive secondary) will result in a wider gap. 

We estimate the accretion of the binary BHs is a typical gap case only when the derived gap width follows the criteria of $W_{\rm gap} \lesssim 1.5 a_{\rm sep}$. For a standard thin disk with a typical aspect ratio of $h = 3 \times 10^{-3}$ (cf. Sec~\ref{sec:jointfitting}) and $\alpha_{\rm vis}=0.3$, a gap width of $W_{\rm gap} = 1.5a_{\rm sep}$ corresponds to the maximum mass ratio for the gap case of $q_{\rm max,gap} \sim 1.6\times10^{-3}$, which is adopted as the upper limit of the gap case in this work. This is a conservative estimate, as simulations of Slim disks (cf. Sec.~\ref{three_models} below) indicate that the gap case can persist up to higher mass ratios, $q \sim 10^{-2}$ (e.g., \citealt{Fung2014}). We further note that, although recent global simulations suggest that turbulence at higher viscosity can generate stochastic torques \citep{Wu-etal.2024,Chen-Wu-etal.2025,Kubli-etal.2026,CevallosSoto-Zhu2026}, we do not take this effect into account here for the sake of simplicity in our analysis.

\subsubsection{CBD case at $q\ga0.04$}

For accretion of binary BHs with a larger mass ratio $q$, the gap width $W_{\rm gap}$ calculated from Eq.~(\ref{eq:gapwidth}) can be wide enough thus the two BHs will clear a broad and deep cavity around their binary orbit. We classify such system as a circumbinary disk case (``CBD case'' hereafter, see \citealt{Munoz2020} for a recent review). The balance of tidal and viscous torques determines the location of the inner edge of the CBD at $r\sim 2a_{\rm sep}$. Numerical simulations indicate that a CBD can form when the mass ratio $q \ga q_{\rm min,CBD}$, where $q_{\rm min,CBD}\sim 0.04$ (\citealt{DOrazio2016}). For intermediate mass ratios between the gap-opening and CBD regimes (i.e., the transitional state), we assume the gap width to be approximately $W_{\rm gap} \sim 2a_{\rm sep}$, adopting the same value as in the CBD case. 

This fundamental configuration, extensively studied for decades, has seen significant advancements recently (see, e.g., \citealt{Lai2023} and references therein). As shown in the right panel of Fig.~\ref{fig:bbh_schem}, based on progress in both theoretical analysis and numerical simulations (e.g., \citealt{Artymowicz1994, Hayasaki2008, Noble2012, Farris2015, Munoz2020}), we model the system as comprising a circumbinary disk (``\texttt{cb-disk}'', green ring), and two mini disks, surrounding the primary (``\texttt{pri-disk}'', blue ring) and secondary (``\texttt{sec-disk}'', purple ring) BHs, respectively. The three accretion disks are connected by gas streams (see, e.g., \citealt{Farris2014, Munoz2020} for numerical simulations), as illustrated by arrows in the figure.

Similar to the secondary disk (\texttt{sec-disk}) in the gap case, the outer truncation radii of the two mini-disks is assumed to be $R_{\rm out} \approx 0.4\,R_{\rm Roche}$. The inner radius of the circumbinary disk (\texttt{cb-disk}) is established at $\sim 2 a_{\rm sep}$, which can also be regarded as the outer edge of the cavity (\citealt{Roedig2014}). For simplicity, in our calculations we neglect the density enhancement of the spiral waves in the circumbinary disks, as well as the radiative contribution from these streams.

\begin{figure}
\centering
\includegraphics[width=0.48\textwidth]
    {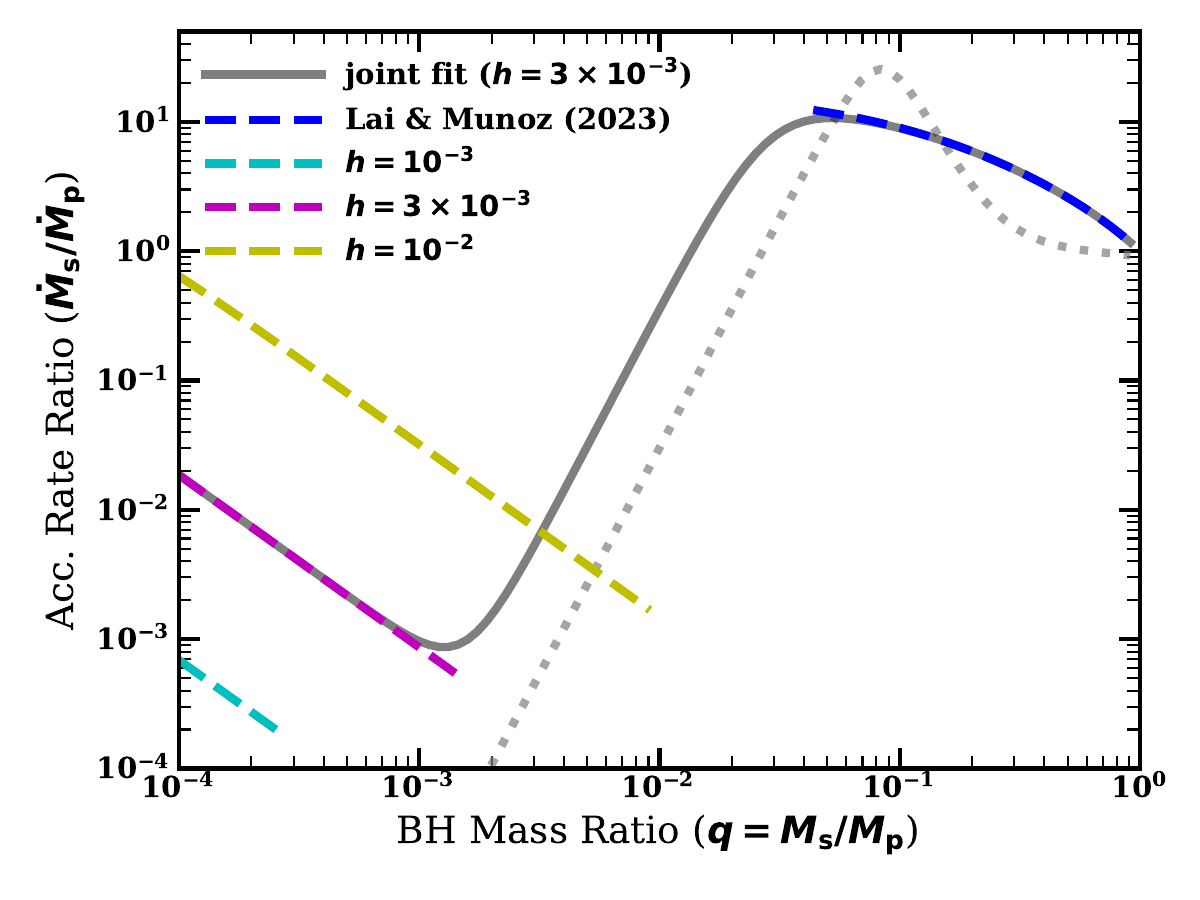}\\
  \includegraphics[width=0.48\textwidth]{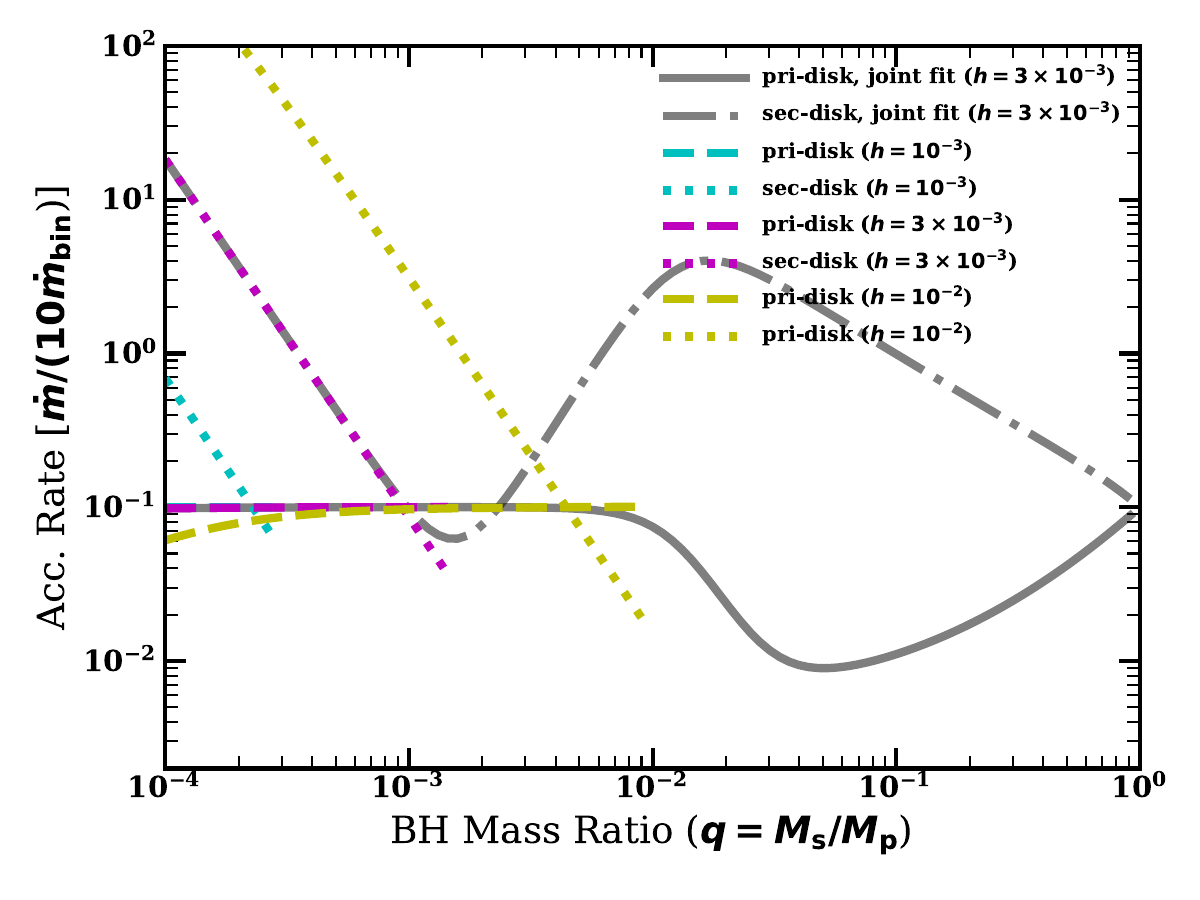}

  \caption{{\it Top panel}: Absolute accretion-rate ratio $\dot{M}_{\rm s}/\dot{M}_{\rm p}$ as a function of the mass ratio $q$. The black dashed line shows the result from \citet{Lai2023}. The cyan, magenta, and yellow dashed curves correspond to the models of \citet{Li2023} for different disk aspect ratios $h
 = 10^{-3}$ (cyan), $3\times10^{-3}$ (magenta) and 0.01 (yellow), respectively. The gray solid and dotted curves show, respectively, the empirical fitting result of this work and \citet{Kelley2019}. Clearly, the secondary received more accreting gas, especially in the CBD case.
  {\it Bottom panel}: Accretion rates of the primary (gray dot-dashed) and the secondary (gray solid) mini-disks, as a function of the binary mass ratio $q$. Dotted and dashed colored lines show accretion rate of the \texttt{pri-disk} and the \texttt{sec-disk} from \citet{Li2023} model for different disk aspect ratios: $h=10^{-3}$ (cyan), $3\times10^{-3}$ (magenta) and $0.01$ (yellow), respectively.}
  \label{fig:bbh_acc}
\end{figure}

\subsection{Binary Massive BH Accretion: Redistribution of Mass Accretion Rate Between Two BHs}

The outer circumbinary disk continues to deliver accreted material and angular momentum through the gap/cavity. An important question of accretion onto binary massive BHs is, given the presence of a gap or a cavity, what fraction of the supplied mass accretion rate, $\dot{M}_{\rm bin}$, is ultimately accreted by each BH?
The ratio of mass accretion rates is crucial for understanding the co-evolution of the binary and the accretion disk system.
This accretion process is highly nonlinear and therefore is primarily studied through multidimensional hydrodynamical simulations. Numerous efforts have been devoted in this field in the past decade for different binary orbital configurations and disk properties (e.g., \citealt{Shi2012, Gold2014, Moody2019, Munoz2020, Li2021b, Duffell2024, Smallwood2025, Wang2025}; see a review by \citealt{Lai2023} and references therein). A similar issue arises for accretion onto extreme-mass-ratio systems as in the case of gas accretion onto an embedded planet (e.g., \citealt{DAngelo2003, Machida2010, Rosenthal2020, Li2021, Li2023, Choksi2023}).

\subsubsection{Gap case}
One important difference to the conventional picture is that, for the accretion of the secondary BH, the Bondi radius may be greater than both the disk height and the Hill radius, and the accretion process will be regulated by the latter two scales \citep{Li2023}. Considering this, we apply the Hill accretion formula from \citet[][]{Li2023} to the extreme-mass-ratio binaries (i.e., ``gap case'') in the AGN disk when $3h^{3}\lesssim q\lesssim q_{\rm max, gap}$, where the effective secondary accretion rate reads,
\begin{equation}
\dot{M}_{\mathrm{s}} = \frac{\dot{M}_{\mathrm{s}}'\dot{M}_{\mathrm{bin}}}{\dot{M}_{\mathrm{s}}' + \dot{M}_{\mathrm{bin}}}.
\label{eq:mdot_ms}
\end{equation}
Here,
\begin{equation}
\dot{M}_{\mathrm{s}}' = \frac{\sqrt{2/9\pi}}{\left(\alpha_{\rm vis} + 0.04 h q_{\mathrm{th}}^2\right)} \left(\frac{q_{\mathrm{th}}}{3}\right)^{2/3} \dot{M}_{\rm bin}, \label{gapacc}
\end{equation}
where $q_{\rm th} \equiv R_{\rm Bondi}/H = q/h^{3}$, and the total accretion rate is defined as $\dot{M}_{\rm bin} \equiv \dot{M}_{\rm p}+\dot{M}_{\rm s}$. The above lower boundary of $q$ ensures that $q_{\rm th}\gtrsim 3$.

We plot in the top panel of Fig.~\ref{fig:bbh_acc} the accretion rate ratio $\lambda(q)\equiv \dot{M}_{\rm s}/\dot{M}_{\rm p}$. Here, three different disk aspect ratios are considered, i.e., $h = 10^{-3}$, $3\times10^{-3}$, and $0.01$. Based on the condition of the gap width $W_{\rm gap} = 1.5 a_{\rm sep}$, the upper limit of mass ratio, $q_{\rm max,gap}$, of the gap case is derived separately for three different choices of $h$.

\subsubsection{CBD case}

For CBD case with $q\ga 0.04$, \citet{Kelley2019} provided a fitting formula for $\lambda(q)$ based on the hydrodynamic simulations of binary BH accretion disks performed by \citet{Farris2014}. In that work, the accretion-rate profile is modeled using a power law with an exponential cutoff (see Eq.~(1) and Fig.~1 in \citealt{Kelley2019}).
Here, we use the most updated fitting result provided by \citet{Lai2023} as shown in Fig.~\ref{fig:bbh_acc}, based mainly on simulations of \citet{Munoz2020} and \citet{Duffell2020}:
\begin{equation}
\frac{\dot{M}_{\rm s}}{\dot{M}_{\rm bin}} \approx 0.5 + \frac{4(1-q)}{9}.
\end{equation}

It should be noted that those simulations usually adopted a large disk aspect ratio compared with the typical value of $H/R\sim 10^{-3}$ for AGN disks \citep{Sirko2003}. The ratio of the mass accretion rate could be different for disks with different disk aspect ratios.

\subsubsection{Joint fit of $\lambda(q)$ $(\equiv\dot{M}_{\rm s}/\dot{M}_{\rm p})$} \label{sec:jointfitting}

Regarding the accretion rate distribution in the transitional regime between the gap and CBD cases, numerical simulations remain limited. To address this, we adopt a fitting approach that bridges these two regimes, allowing us to derive a unified fitting formula relating the mass ratio $q$ to $\lambda(q)$ ($\equiv\dot{M}_{\rm s}/\dot{M}_{\rm p}$) over a much broader range of $q = 10^{-4}$ to 1.

Following the spirit of \citet{Kelley2019}, we use an exponential function to fit the rapid change in $\lambda(q)$ at the intermediate mass ratios, which reads,
\begin{eqnarray}\label{eq:fitting}
 \lambda_{\rm fit}(q) & = & \frac{\lambda_{\rm gap}(q) \cdot \lambda_{\rm CBD}(q)}{w \cdot \lambda_{\rm CBD}(q) + (1 - w) \cdot \lambda_{\rm gap}(q)}, \\
 w & = & \frac{1}{1 + \exp\!\big[k\,(0.5 - t)\big]}, \hspace{0.3cm} t =  \frac{\log_{10}(q/q_2)}{\log_{10}(q_1/q_2)}.\nonumber
\end{eqnarray}
Here $\lambda_{\rm gap}$ and $\lambda_{\rm CBD}$ correspond to the gap and the CBD cases, respectively. $w(t)$ is a weight function of $t$.
$q_1$, $q_2$, and $k$ are fitting parameters. To make a smooth transition for the $\lambda$ profile from the CBD to the gap case (for the $h=3\times10^{-3}$ case), we find $q_1 = 0.012$, $q_2 = 0.0002$, and $k = 20$. The fitting result is shown by the grey solid curve in Fig.~\ref{fig:bbh_acc}. 

The accretion rates presented in Fig.~\ref{fig:bbh_acc} are derived from this fitting formulae. Specifically, the accretion rate ratio in the Eddington unit of each BH have: 
$\dot{m}_{\rm p/s} = \dot{M}_{\rm p/s}/{\dot{M}_{\rm {Edd, p/s}}}$. 
Note that this analysis is performed for a disk aspect ratio $h = 3\times 10^{-3}$, which is consistent with that of a cold thin circumbinary disk at $10^{2-4} R_{\rm g}$ whose binary mass $M_{\rm bin}=10^8\msun$, accretion rate $\dot{m}_{\rm bin}=0.1$ and viscosity parameter $\alpha_{\rm vis}=0.3$. The qualitative behavior remains similar for other values of $h$. Considering that $\dot{m}_{\rm p/s}$ depends linearly on $\dot{m}_{\rm bin}$, the bottom panel of Fig.~\ref{fig:bbh_acc} shows  $\dot{m}_{\rm p/s}/(10\dot{m}_{\rm bin})$.

In the gap region of $q\lesssim q_{\rm max, gap}$, as the mass ratio $q$ increases, the gas density within the gap decreases, leading to a decrease in $\dot{M}_{\rm s}$ correspondingly \citep{Li2023}. As $q$ exceeds its maximal value $q_{\rm max, gap}$, the system transitions to an intermediate state between the gap and the CBD cases. In the gap region of $10^{-3}\lesssim q\lesssim 2.5\times10^{-3}$, $\dot{m}_{\rm s}$ falls below $\dot{m}_{\rm p}$. Subsequently, in the transition region $\dot{m}_{\rm s}$ begins to rise until $q\sim q_{\rm min, CBD}\simeq0.04$, after which the system fully enters into the CBD-dominated accretion regime.

\subsection{Three Accretion Models Depending on Accretion Rate: SSD, HAF, and Slim Disk}\label{three_models}

There are in general three accretion states, depending primarily on the mass accretion rate (and magnetic flux) of the systems. In this section, we will provide a brief description of them. We note that most previous studies of binary massive BH accretion have focused on the cold Shakura-Sunyaev disk (SSD, \citealt{Shakura1973, Novikov1973}). One exception is the recent work by \citet{Tiede2025}, which explored accretion models for sub-Eddington accretion in binary SMBH systems, with a particular emphasis on the spectral properties of hot accretion flow (HAF, \citealt{Yuan2014}. Originally named Advection-Dominated Accretion Flow/ADAF, \citealt{Narayan1994}). Their research systematically explores the spectral manifestations of multi-component accretion flows, where solutions can transit independently between cold SSD and hot HAF configurations based on the local accretion rate. For the case we study with $\dot{m}_{\rm bin}=0.1$ (and $h\sim3\times10^{-3}$), the accretion rate of secondary mini-disk can rise to $\dot{m}_{\rm s}\sim 10-100$ at $q\lesssim10^{-4}$ (see Fig.~\ref{fig:bbh_acc}). Under such conditions, the Slim disk solution must be adopted for the \texttt{sec-disk}. As shown below, the luminosity of the \texttt{sec-disk} is effectively suppressed when it is a Slim disk.

The widely adopted SSD model describes an optically thick but geometrically thin ($h\ll 1$) accretion disk \citep{Shakura1973, Novikov1973}. In this framework, nearly half of the gravitational potential energy released by infalling matter is converted into radiation. Each annular ring of the disk approximately reaches a local thermal equilibrium and radiates as a blackbody. Matter spirals inward on near-Keplerian orbits, with a temperature profile $T \propto R^{-3/4}$ \citep{Shakura1973}. Consequently, the inner regions attain higher temperatures and emit primarily in the ultraviolet （and soft X-rays if BH mass is small) band, while the cooler outer regions radiate predominantly in the optical and infrared bands. Depending on BH spin, the radiative efficiency $\epsilon = L_{\rm bol}/\dot{M} c^2$ (here $L_{\rm bol}$ is the bolometric luminosity) is $5.7-42\%$ with a typical value of $\sim 10\%$ \citep{Novikov1973}. The SSD model has successfully applied to bright AGNs and soft state of black hole X-ray binaries \citep{Frank2002}. 

The HAF model \citep{Narayan1994} represents a crucial extension of the SSD paradigm at low $\dot{m}$ regime, where it becomes tenuous and optically thin. As a result, the viscous/turbulent heating is not radiated locally but instead is stored as thermal enthalpy and advected inward with the accreting gas across the BH horizon. This mechanism leads to a high ion temperature $T_{\rm i}\sim10^{10}$K, thus HAF is geometrically thick ($H/R\sim 1$). Radiative cooling to electrons keeps $T_{\rm e}$ much lower than $T_{\rm i}$ \citep{Xie2012}. Consequently, HAF consists two-temperature plasma \citep{Narayan1995, Mahadevan1997}. The emission of HAF is predominantly non-thermal \citep{Narayan1995}, i.e., a radio-infrared produced via synchrotron and a optical-(hard)X-rays continuum produced via inverse Compton scattering. Bremsstrahlung emitted by hot gas at large radius also contributes in X-rays.

There are several important updates in HAF (cf. \citealt{Xie2012,Yuan2014} for a summary). At accretion rate larger than that of typical HAF, the accretion flow can remain hot, thanks to the compression work of the accreting gas. In this regime, the radiative efficiency increases dramatically. This is because of strong Compton scattering that has a non-linear steep dependence on optical depth, provided the optical depth is moderately large. At even higher accretion rate regime, the accretion flow will turn into two-phase, i.e., numerous cold clumps are formed due to strong radiative cooling, and they are embedded in hot gases \citep{Yang2015, Wu2016}. In this case, radiative efficiency  can reach a value as high as $\sim 8\%$, making it comparable to that of SSD \citep{Xie2012}. HAF has been applied to understand low-luminosity AGN and black hole X-ray binaries in their hard and intermediate states (for a review, see e.g., \citealt{Yuan2014}). 

In this work, the HAF model is from \citet{Xie2012}. Following their study, the fraction of viscous heating that directly heat the electrons is set to $\delta=0.1$. For this $\delta$, the critical accretion rate of HAF is given by $\dot{m}_{\rm crit,HAF}={\rm max}(0.5 \alpha_{\rm vis}^2, 0.066 \alpha_{\rm vis})$. For our chosen value of $\alpha_{\rm vis}$, we have $\dot{m}_{\rm crit,HAF}= 0.045$. We note that the two-phase regime is highly suppressed for a large $\alpha_{\rm vis}$, and in this work we omitted it for simplicity. In our calculations of HAF, considering new progress in accretion theory \citep{Narayan2003}, we take a HAF with stronger magnetic field strength, i.e. the gas-to-magnetic pressure ratio is set to $\beta = 0.5$. Furthermore, outflow is taken into account by setting $\dot{M}(R)=\dot{M}_{\rm out}(\frac{R}{R_{\rm out}})^{0.3}$, where $R_{\rm out}$ is outer boundary of each accretion disk. For the two mini-disks, $\dot{M}_{\rm out}$ is directly calculated from Eq.~(\ref{eq:fitting}). 

The Slim disk represents a distinct accretion regime that occurs at high accretion rates approaching or exceeding the Eddington limit \citep{Abramowicz1988}. It is invoked to explain many highly luminous systems, such as ultraluminous X‑ray sources (e.g., \citealt{King2001}) and narrow-line Seyfert 1 galaxies(e.g., \citealt{Collin2004}). In this luminosity regime, radiation pressure dominates the total pressure—in contrast to the gas-pressure dominance in SSD—leading to a geometrically thickened structure with moderate aspect ratio as high as $h\sim 0.1-0.5$ \citep{Abramowicz1988}. Radiative cooling is suppressed as most photons are absorbed and scattered within the dense gas, a process referred to as ``photon trapping'' \citep{Abramowicz1988, Ohsuga2002}. Consequently, radiation efficiency of Slim drops much below $\sim10\%$, see Fig.~\ref{fig:efficiency} below). In addition, the radial infall velocity of Slim approaches is comparable to the free‑fall speed (e.g., $v_r\sim 0.1c$), which is significantly faster than that of SSD.

For $\dot{m}>\dot{m}_{\rm crit,HAF}$, we adopt \citet{Kubota2019} that has a joint description of both SSD and Slim disk under general relativistic (GR) frame (\citealt{Novikov1973}, NT hereafter). The critical accretion rate $\dot{m}_{\rm crit,SSD}$ demarcates the transition between the regimes of SSD and Slim: the Slim disk becomes physically relevant at accretion rates $\dot{m}_{\rm crit,SSD}\sim 5-6$.\footnote{Note that a different definition of Eddington accretion rate is adopted in \citet{Kubota2019}, i.e. $\dot{M}_{\rm Edd} = L_{\rm Edd} /(\epsilon c^2)$, where radiative efficiency $\epsilon$ is a function of accretion rate \citep{Abramowicz1988} and BH spin. We additionally caution that, observers usually adopt another definition, $\dot{M}_{\rm Edd} = L_{\rm Edd}/c^2$, which excludes the 10\% efficiency thus is ten times smaller than our choice.} When $\dot{m} < \dot{m}_{\rm crit,SSD}$, the SSD radiative flux $F_{\rm SSD} (R)$ is lower than the static, spherically symmetric local Eddington flux $F_{\rm Edd}(R)=L_{\rm Edd} /(4\pi R^2)$, and the accretion flow retains a canonical SSD structure. For $\dot{m} > \dot{m}_{\rm crit,SSD}$, the NT-model-computed flux $F_{\rm SSD}(R)$ exceeds the local Eddington flux $F_{\rm Edd}(R)$ in the radial range $R_{\rm bc} < R < R_{\rm crit}$ (see Fig.~1 of \citealt{Kubota2019}). In this region, the local Eddington flux acts as an upper limit: the intrinsic emissivity of the inner disk is suppressed relative to that of a SSD (see Fig.~\ref{fig:efficiency}), such that the local disc flux reduced to its saturated value, i.e. $F_{\rm Slim}(R)=F_{\rm Edd}(R)$ (see Eq.~2 in \citealt{Kubota2019}). The effective temperature of Slim disk in the $R_{\rm bc} < R < R_{\rm crit}$ region can then be derived by $\sigma_{\rm SB} T_{\rm eff,Slim}^4(R) = F_{\rm Slim}(R)$, i.e. it recovers the shallower temperature profile $T_{\rm eff,Slim}\propto R^{-1/2}$ \citep{Abramowicz1988}.

\begin{figure}
\centering
  \includegraphics[width=0.4\textwidth]{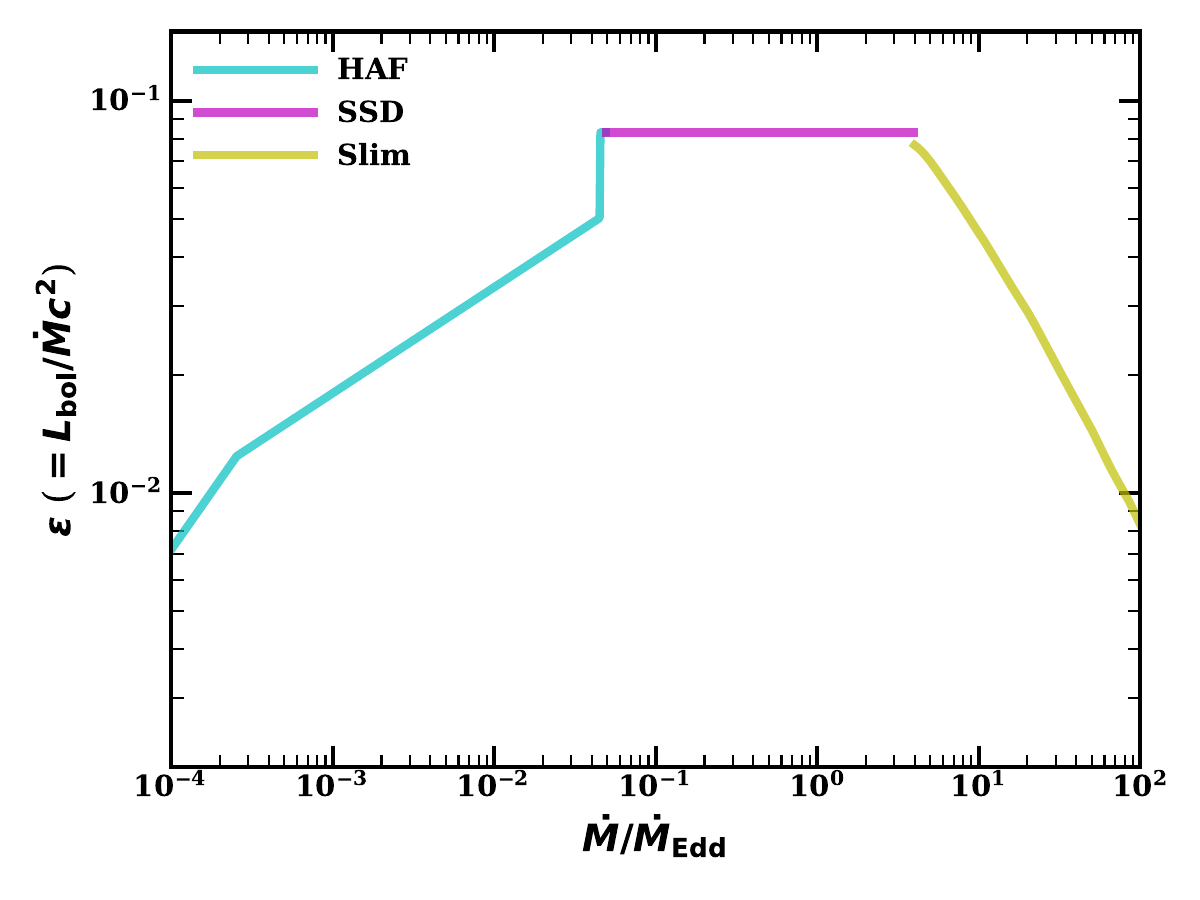}
  \caption{Radiative efficiency $\epsilon$ (defined as $L_{\rm bol}/\dot{M}c^{2}$) as a function of accretion rate in Eddington unit $\dot{m} = \dot{M}/\dot{M}_{\rm Edd}$ (Note that $\dot{M}_{\rm Edd} = 10L_{\rm Edd}/c^2$ is adopted). Depending on $\dot{m}$, there are three accretion regimes, i.e. HAF (cyan), SSD (magenta), and Slim (yellow). The steep increase in $\epsilon$ at the bright end of the HAF regime is because of strong Compton scattering effect when the optical depth is high \citep{Xie2012}. In this calculation, the black hole spin is set to $a^{\star}=0.5$.}
  \label{fig:efficiency}
\end{figure}

Figure ~\ref{fig:efficiency} summarizes the radiative efficiency of all three accretion models. The magneta and yellow curves are the radiative efficiency of SSD and Slim disk models, respectively, following \citet{Kubota2019}. Obviously, the radiative efficiency of Slim disk decreases with increasing accretion rate \citep{Watarai2000}. The cyan curve shows the radiative efficiency of HAF, where we take the fitting formulae from \citet{Xie2012}.

\section{Radiative Properties of Binary Massive BH Accretion}\label{sec:rad}

\subsection{Luminosity differences among two BHs at Different Mass Ratios $q$}

Before detailed calculations of outcome spectral energy distribution (SED), we first provide a crude estimation of luminosities of each mini-disk. 

\begin{figure}
\centering
   \includegraphics[width=0.45\textwidth]{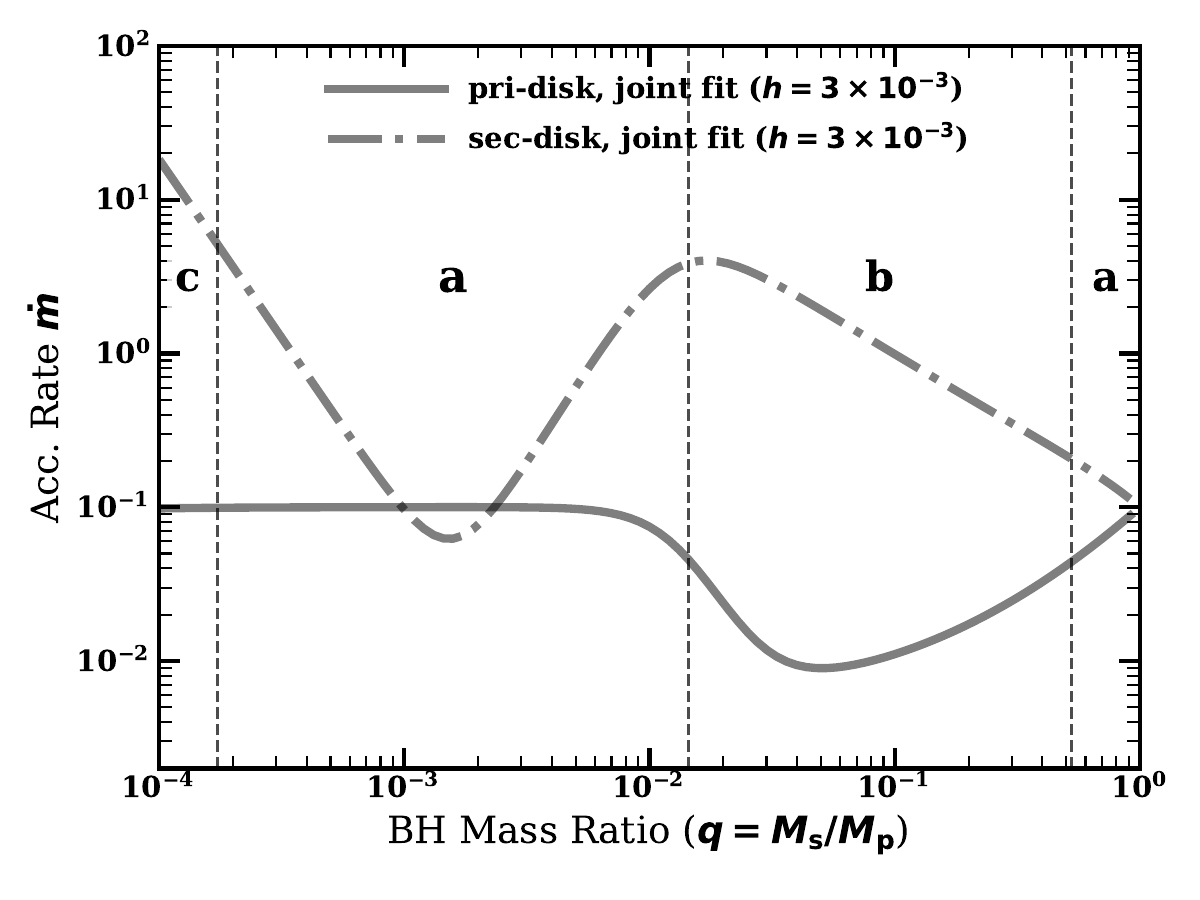}
  \caption{Accretion rates (in their respective Eddington unit) of the mini-disks around the primary (gray dot-dashed curve) and the secondary (gray solid) BH, as a function of BH mass ratio $q$ (based on joint fitting of $\lambda(q)$ with $h=3\times10^{-3}$). The total accretion rate is $\dot{m}_{\rm bin}=0.1$. Regions a, b, and c correspond to the configurations  of ``SSD + SSD'', ``HAF + SSD'', and ``SSD + Slim'' (\texttt{pri-disk} + \texttt{sec-disk}), respectively.}
  \label{fig:acc_disktypes}
\end{figure}

For our adopted total accretion rate $\dot{m}_{\rm bin}=0.1$, based on its redistribution among two BHs, we can classify the disk around the primary and secondary BHs into different accretion models. This is shown in Fig.~\ref{fig:acc_disktypes}, where . The \texttt{cb-disk} is always an SSD, and it not shown here for simplicity. According to different accretion modes, we separate into three regions, i.e. Regions a, b, and c, to mark the configurations of ``SSD + SSD'', ``HAF + SSD'', and ``SSD + Slim'' (\texttt{pri-disk} + \texttt{sec-disk}), respectively.

\begin{figure}
\centering
\includegraphics[width=0.45\textwidth]{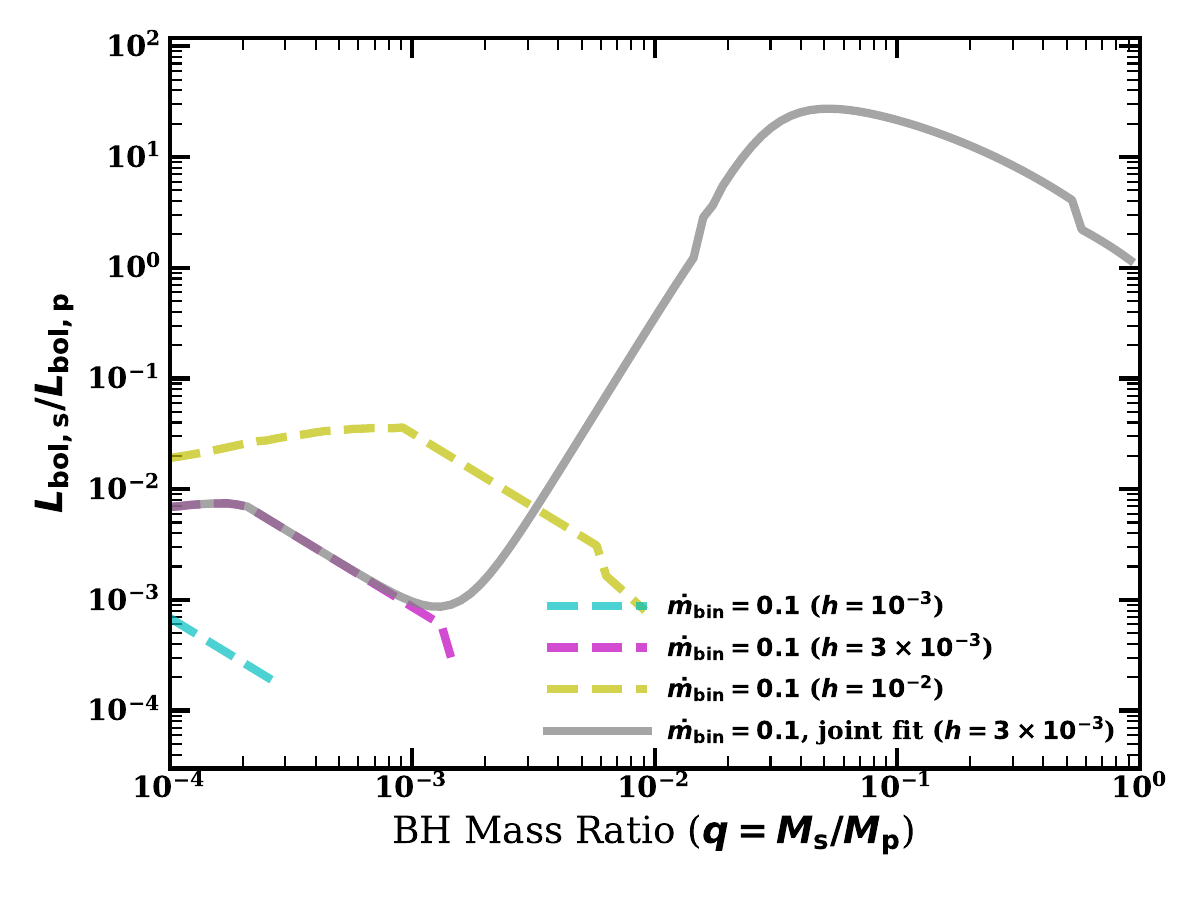}
  \caption{Bolometric luminosity ratio $L_{\rm bol,s}/L_{\rm bol,p}$ of the secondary to primary mini-disks as a function of the mass ratio. The gray solid curve represents out fitting. The cyan, magenta, and yellow dash curves correspond to models with disk aspect ratio $h=10^{-3}, 3\times10^{-3}$ and $0.01$, respectively. The black hole spin is set to $a^{\star}=0.5$ and the total binary accretion rate is $\dot{m}_{\rm bin} = 0.1$.}
  \label{fig:lbol_q}
\end{figure}

Using the radiative efficiency as a function of accretion rate (Fig.~\ref{fig:efficiency}), we can then derive, for these different disk configurations, the secondary-to-primary luminosity ratio $L_{\rm bol,s}/L_{\rm bol,p}$ for different mass ratios, and the results are shown in Figure~\ref{fig:lbol_q}. Supplement model results of the gap case \citep{Li2023} with $h=10^{-3}, 3\times10^{-3}, 10^{-2}$ are also shown here. From this figure, we have two clear observations. First, in the CBD case, because the secondary BH received more accreting gas, most emission of this binary SMBH system actually comes from \texttt{sec-disk}. Second, in the gap case (and the transition regime), emission from \texttt{sec-disk} is highly reduced, $L_{\rm bol,s}/L_{\rm bol,p}\sim 10^{-3} - 10^{-2}$ if $q<3\times10^{-3}$. The sharp drop in $L_{\rm bol,s}/L_{\rm bol,p}$ is a consequence of the combination of differences in $\dot{M}_{\rm s}/\dot{M}_{\rm p}$ (cf. Fig.~\ref{fig:bbh_acc}) and radiative efficiency (cf. Fig.~\ref{fig:efficiency}) among different accretion models.

\begin{deluxetable}{lcccccl}
\tablecaption{Summary of Model Parameters and Configurations\label{tab:model_par}}
\tablewidth{0pt}  %
\tablehead{
\colhead{Parameter} & \colhead{Model 1} & \colhead{Model 2} & \colhead{Model 3} & \colhead{Model 4} &
\colhead{Model 5} &
\colhead{Description} \\
\colhead{} & \colhead{(Fig.~\ref{fig:q=0.5})} & \colhead{(Fig.~\ref{fig:SED_q=0.1})} & \colhead{(Fig.~\ref{fig:SED_three} left)} & \colhead{(Fig.~\ref{fig:SED_three} middle)} & \colhead{(Fig.~\ref{fig:SED_three} right)} &
}
\startdata
Total BH mass ($M_{\rm bin}$) & \multicolumn{5}{c}{$10^8\,M_\odot$ (fixed)} & Total mass of binary massive BHs \\
Total BH accretion rate ($\dot{m}_{\rm bin}$) &  \multicolumn{5}{c}{$0.1$ (fixed)} & $ \dot{M}/\dot{M}_{\rm Edd,bin}$\\
Orbital eccentricity ($e$) & \multicolumn{5}{c}{0 (fixed)} & Orbital eccentricity of binary BHs\\
BH spin ($a^\star$) & \multicolumn{5}{c}{0.5 (fixed)} & dimensionless spin parameter\\
\hline
Mass ratio ($q$) & $0.5$ & 0.1 & $10^{-2}$ & $10^{-3}$ & $10^{-4}$ & $M_{\rm s}/M_{\rm p}$, BH mass ratio \\
Binary separation ($a_{\rm sep}$) & $50-5\times10^4$& $2\times10^3$& $2\times10^3$& $2\times10^3$& $2\times10^3$& Separation of two BHs, in unit $GM_{\rm bin}/c^2$ \\
Binary state & CBD & CBD & transition & gap & gap & Configuration of binary massive BH accretion\\
Accretion model & SSD+SSD & HAF+SSD & SSD+SSD & SSD+SSD & SSD+Slim & Accretion of \texttt{pri-disk} + \texttt{sec-disk} \\
\hline
Separation ($a_{\rm sep}$) for $t_{\rm gw} = 100$ yr  & 90 & 73 & 43 & 24 & 13 & Only for panels in Fig.~\ref{fig:meger_two} \\
\enddata
\tablenotetext{}{The separations are for panels in Fig.~\ref{fig:meger_two}.}
\end{deluxetable}

\subsection{Broadband Spectrum}

\begin{figure}
\centering
    \includegraphics[width=0.45\textwidth]{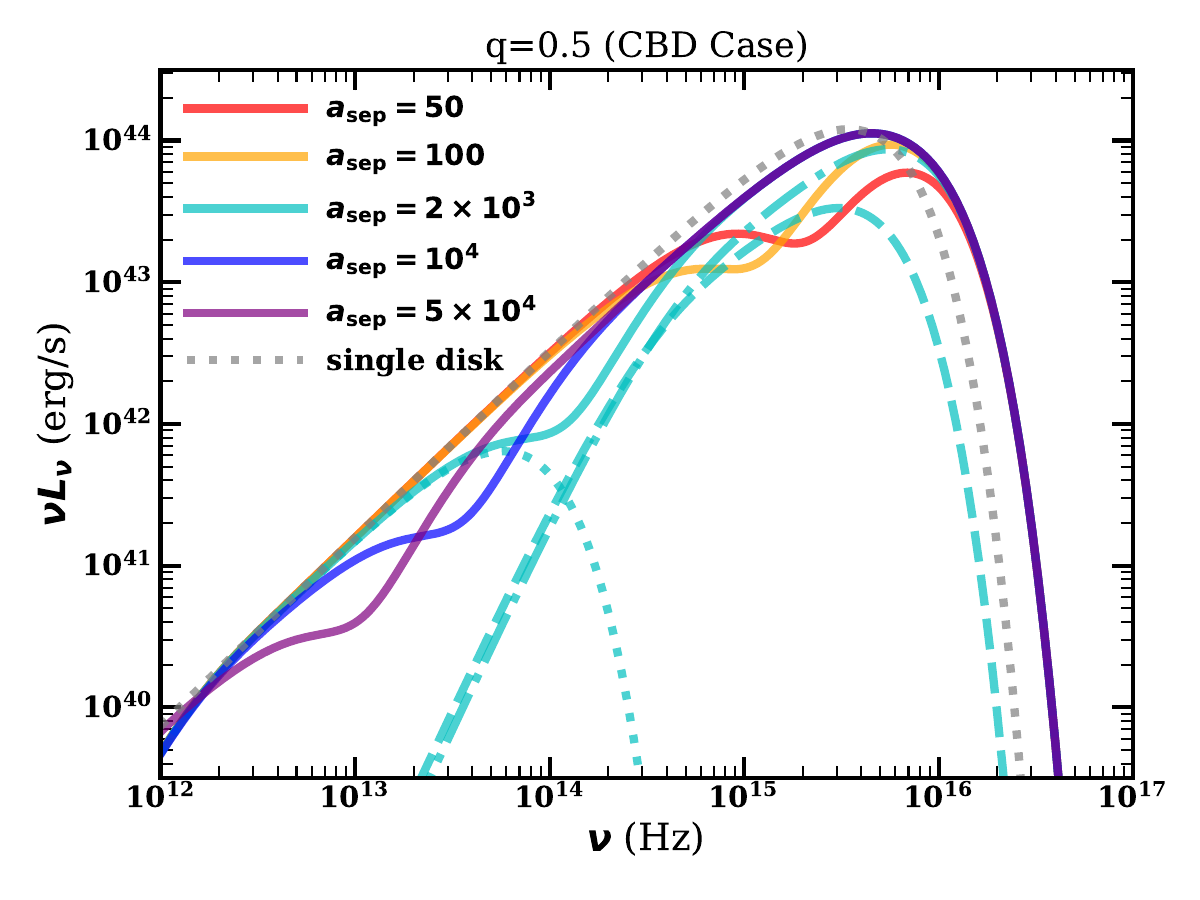}	
    \caption{SEDs for a binary BH system with a mass ratio $q=0.5$.
     The system has a total binary mass $M_{\rm bin}=10^8\msun$ and an accretion rate $\dot{m}_{\rm bin}=0.1$. The plot shows the total SEDs for different orbital separations ($a_{\rm sep}$ = 50, 100, $2\times10^3$, $10^4$, $5\times10^4R_{\rm g}$), represented by solid lines in red, orange, cyan, blue, and purple, respectively. Each SED (colored solid line) is the superposition of three components: \texttt{cb-disk} （dotted), \texttt{pri-disk} (dashed), and \texttt{sec-disk} (dot-dashed), where for clarity only the decomposition for $a_{\rm sep}=2\times10^3 R_{\rm g}$ is illustrated. For comparison, the gray dashed line shows the SED of a corresponding SSD onto a single BH with $\mbh=10^{8}\msun$.}
     \label{fig:q=0.5}
\end{figure}

We first investigate the impact of BH separation $a_{\rm sep}$. In Fig.~\ref{fig:q=0.5}, we show the SEDs from binary BH disks with separations ranging from $50 R_{\rm g}$ to $5\times10^4 R_{\rm g}$ for a fixed mass ratio $q=0.5$. The binary accretion system model corresponds to the CBD case. 
In this case, all three disks are SSDs. The presence of a central cavity in the accretion flow gives rise to a prominent ``notch'' in the spectrum (e.g., \citealt{Sesana2012, Tanaka2012, Tanaka2013, Farris2015, Yan2015, Zheng2016, Inayoshi2026}). The frequency of this notch, denoted as $\nu_{\rm notch}$, corresponds to the peak frequency of the blackbody spectrum emitted by an unperturbed circumbinary disk $\nu_{\rm un, peak}$ at the secondary's orbital radius $R_{\rm sec}$. It can be seen that the notch location $\nu_{\rm notch}$ moves to the higher energy band as $a_{\rm sep}$ decreases. This can be simply understood as follows. According to the SSD model, the far-infrared to optical spectra follow $F_{\nu}\varpropto\nu^\alpha$ with $\alpha=1/3$, a consequence of the effective temperature profile $T_{\rm eff}\varpropto R^{-3/4}$ \citep{Shakura1973} for the circumbinary disk.
Consequently, this implies a scaling relation $\nu_{\rm notch}\propto R_{\rm sec}^{-3/4}$. By fitting the notch location for different $a_{\rm sep}$, it yields (note that $R_{\rm sec}\approx a_{\rm sep}$ if $q\ll 1$),
\begin{equation}
    \nu_{\rm notch}\propto a_{\rm sep}^{-0.76}.
\end{equation}
This aligns with the theoretical expectation derived above.

\begin{figure}
\centering
    \includegraphics[width=0.45\textwidth]{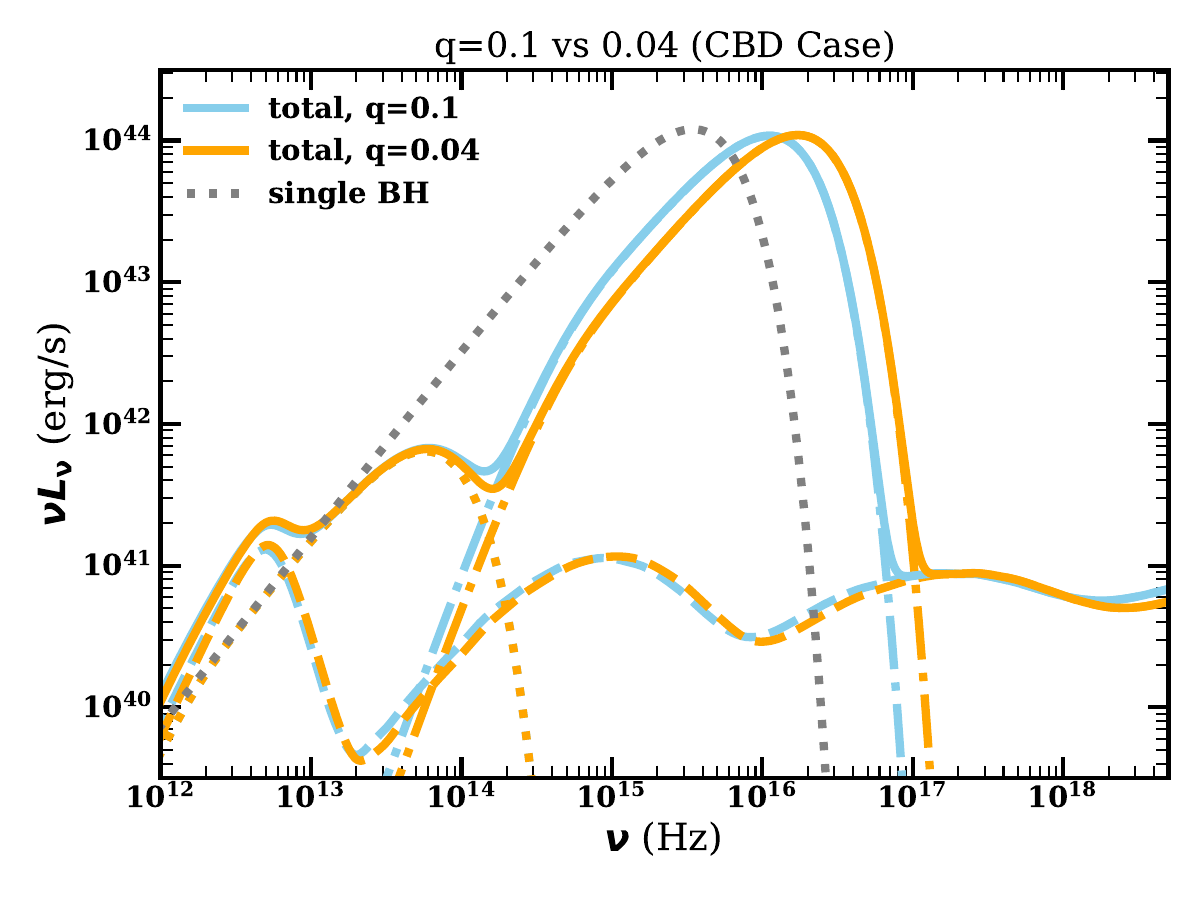}
    \caption{SEDs of CBD case for two binary BH systems with nearly the same accretion rate redistribution $\lambda(q)$. Basic parameters are: $M_{\rm bin}=10^8{\msun}$, $\asep=2000 R_{\rm g}$, and $\dot{m}_{\rm bin}=0.1$. The skyblue and orange lines present SEDs with $q=0.1$ and $q=0.04$, respectively. The line-style conventions used to represent the different disk components are consistent with those in Fig. \ref{fig:q=0.5}.}
    \label{fig:SED_q=0.1}
\end{figure}

Now we investigate the impact of BH mass ratio $q$ on the SED of accretion onto binary BHs. Basic parameters and model configurations are summarized in Table~\ref{tab:model_par}. The separation of two BHs is set to $a_{\rm sep}=2000\ R_{\rm g}$, and we further consider four mass ratios, i.e. $q=0.1, 0.01, 10^{-3}, 10^{-4}$, respectively. We note that, for our adopted $M_{\rm bin}$, the secondary BH becomes an IMBH if $q\lesssim 10^{-2}$.

For $q=0.1$,  which corresponds to a CBD case, the total emission arises from an ``HAF + SSD'' (\texttt{pri-disk} + \texttt{sec-disk}) configuration, as shown in Figure~\ref{fig:SED_q=0.1}. The infrared-to-UV emission is dominated by the ``\texttt{cb-disk}'' and the ``\texttt{sec-disk}'', both of which are SSDs. Most of the accreting material is funneled to the \texttt{sec-disk}, which therefore radiates more strongly than the \texttt{pri-disk}. However, the accretion rate of the \texttt{sec-disk} remains below the critical value of $\dot{m}_{\rm crit,SSD}$ as required for the onset of a Slim disk, thus it remains an SSD. In the outer region of the \texttt{sec-disk}, the effective temperature follows $T_{\rm eff}\varpropto R^{-3/4}$ until the disk is truncated at $0.4R_{\rm Roche}$. 
This truncation produces a marked break in the spectral slope, changing from $\alpha=1/3$ to $\alpha > 1/3$ in the infrared-optical band.
The SED peaks in the far-UV, near $10^{16}$ Hz. Compared to the emission of an SSD around a single BH, the SED from the \texttt{sec-disk} shifts to higher frequencies, a consequence of smaller black hole mass \citep{Shakura1973, Frank2002}. Conversely, the accretion rate onto the \texttt{pri-disk} is so low that it becomes an HAF. The inverse Compton scattering from HAF provides an additional, non-thermal radiation component, extending from the X-ray to the gamma-ray band. The synchrotron bump at $\sim (3-20)\times 10^{12}$ Hz is also evident.

\begin{figure*}
    \centering
    \includegraphics[width=0.3\textwidth]{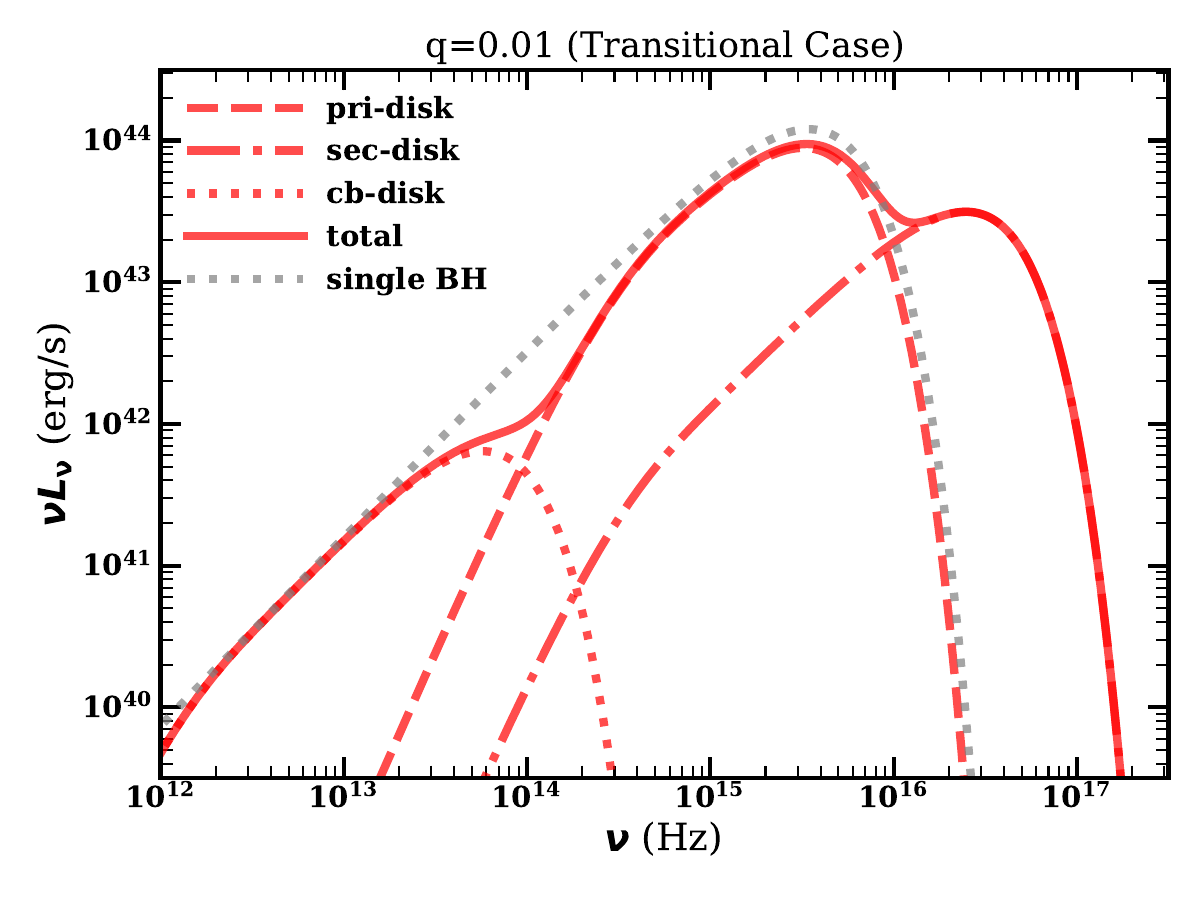}
    \includegraphics[width=0.3\textwidth]{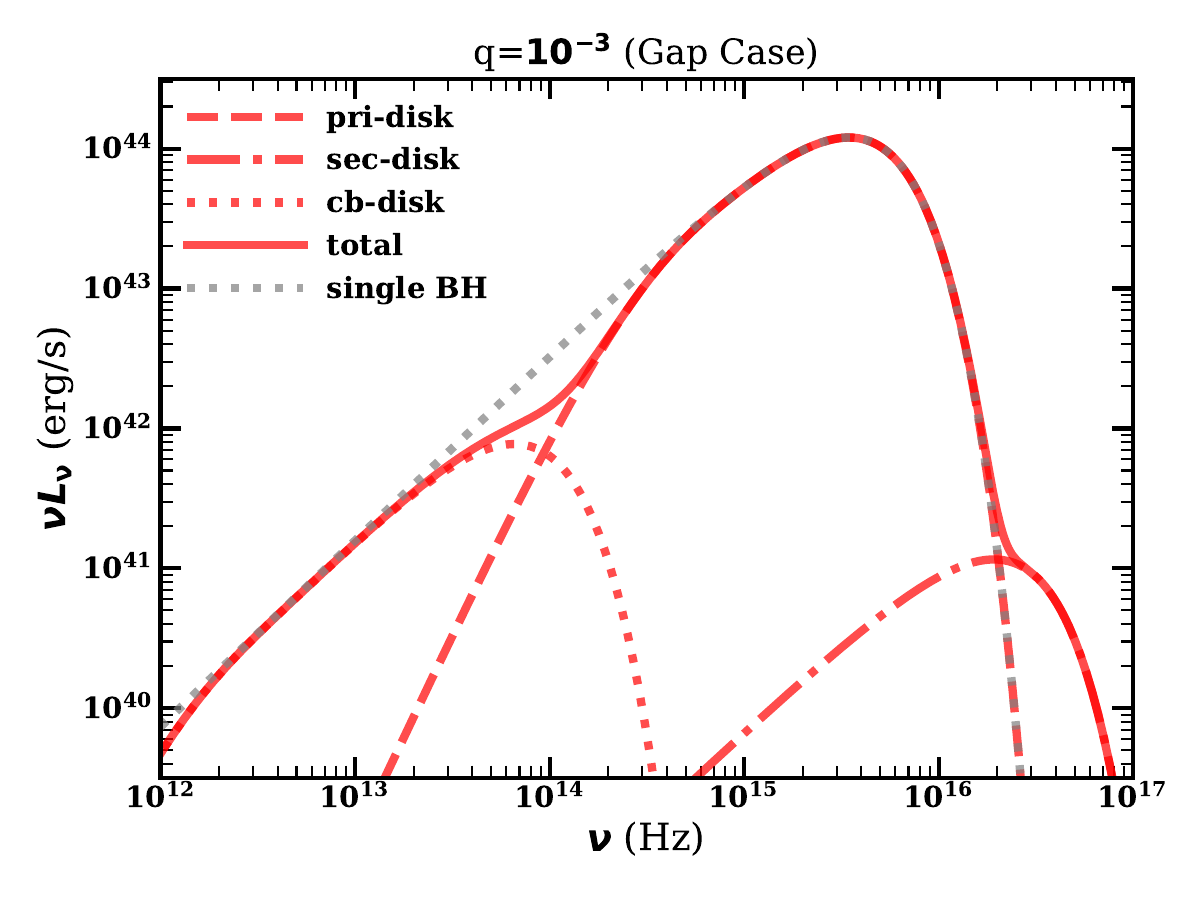}
    \includegraphics[width=0.3\textwidth]{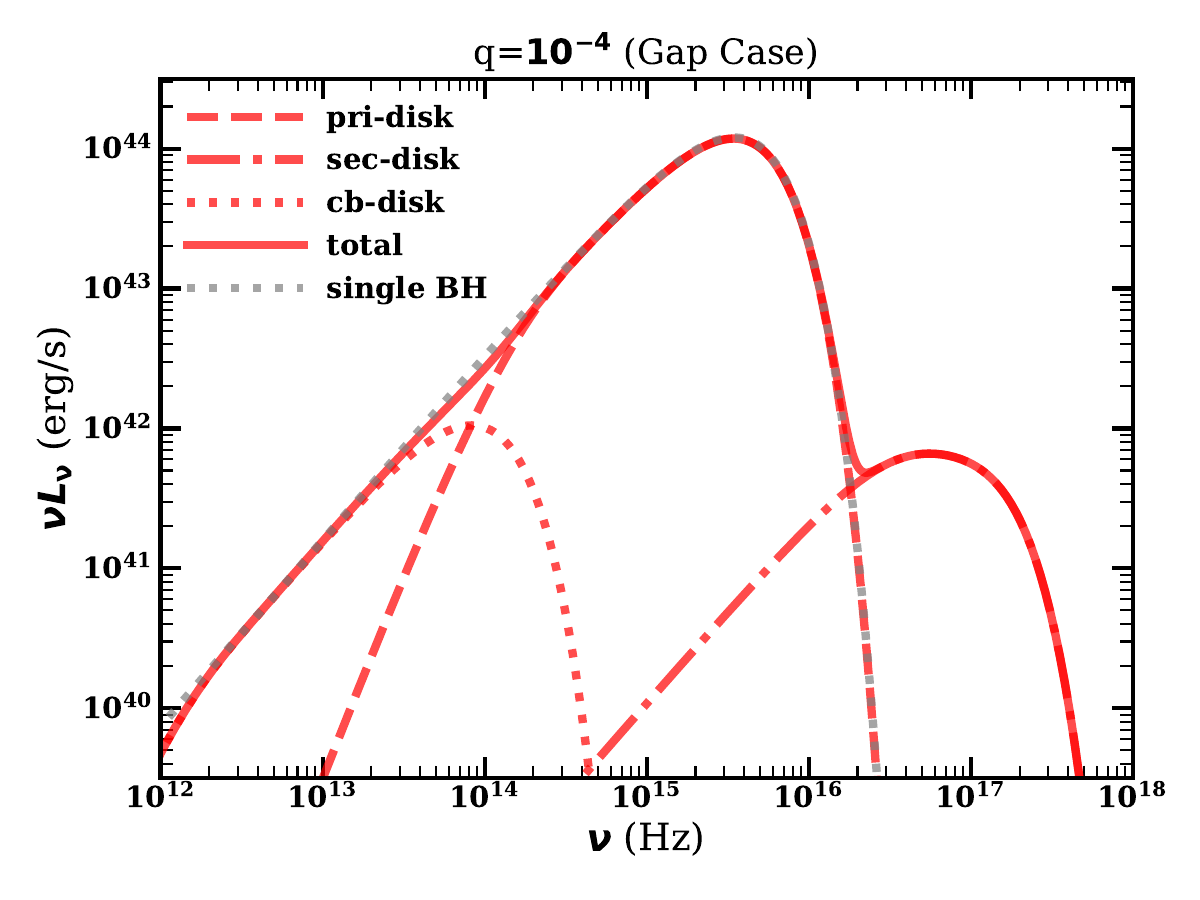}
    \caption{SEDs of binary BH accretion systems with mass ratio $q= 0.01$ ({\it left panel}, transitional case), $q=10^{-3}$ ({\it middle panel}, gap case), $q= 10^{-4}$ ({\it right panel}, gap case), respectively. Each system has a total BH mass $M_{\rm bin}=10^8\msun$, an accretion rate $\dot{m}_{\rm bin}=0.1$, and a binary separation $\asep=2000 R_{\rm g}$. The line-style conventions used to represent the different disk components are consistent with those in Fig. \ref{fig:q=0.5}.}
    \label{fig:SED_three}
\end{figure*}

In the left and middle panels of Figure~\ref{fig:SED_three}, we show the SEDs of radiation from three cold disks at a separation of $2000R_{\rm g}$ for $q=0.01$, $q=10^{-3}$, both of which show ``SSD + SSD'' configurations. For the case of $q=0.01$, the binary enters a transitional regime between the gap and CBD cases. 
In the far-UV, a spectral slope break occurs, marking the transition from \texttt{pri-disk} dominance to \texttt{sec-disk} dominance. However, atmospheric absorption in ultraviolet makes this band challenging to observe. We note that, emission from such systems at high redshift (i.e. $z\sim 5$ or even higher) may be shifted to optical or near-UV band, which can be partially addressed by the next generation of space‑based UV telescopes -- including proposed missions such as the Ultraviolet Explorer (UVEX, \citealt{Kulkarni2021}) and the Habitable Exoplanet Observatory (HabEx, \citealt{Gaudi2018}) -- and broad-band telescopes that includes UV (e.g., the Vera C. Rubin Observatory’s Legacy Survey of Space and Time; LSST, \citealt{Brandt2018}; the Chinese Space Station Survey Telescope; CSST, \citealt{CSST2026}; the Large UV/Optical/IR Surveyor; LUVOIR, \citealt{LUVOIR2019}). 

When the mass ratio reduces to $q=10^{-3}$, the system enters the gap case. The accretion rate of \texttt{sec-disk} drops to sufficiently low values that the \texttt{sec-disk} becomes faint and no longer readily detectable. Nevertheless, the spectral ``notch'' feature, produced by the circumbinary gap/cavity, remains a prominent diagnostic signature.

When the mass ratio decreases to $q=10^{-4}$, the accretion of binary BHs is the gap case. \texttt{sec-disk}, with its accretion rate exceeds $m_{\rm {crit, SSD}}$, transits to a Slim disk, and the system corresponds to a ``SSD+ Slim'' configuration. In this case, \texttt{sec-disk} makes a notable contribution in soft X-rays. However, because the super-Eddington accretion suppresses the temperature in the inner disk region, thus reduces the radiative efficiency of Slim model (cf. Fig~\ref{fig:efficiency}), the total luminosity does not increase indefinitely as $q$ decreases (and $\dot{M}_{\rm s}/\dot{M}_{\rm p}$ increases). Instead, the luminosity growth saturates, remaining roughly two orders of magnitude lower than that of the \texttt{pri-disk}. Moreover, owing to the combination of low black hole mass and high accretion rate, the \texttt{sec-disk} radiates predominantly in the X-ray band. This distinctive signature serves as a strong observational indicator for the presence of a binary massive BH system.

One parameter to characterize the optical-to-X-ray broadband spectrum is through the ``apparent'' spectral index $\alpha_{\rm ox}$ (defined as flux $F_\nu \propto \nu^{-\alpha_{\rm ox}}$), thus $\alpha_{\rm ox}$ can be crudely measured by fluxes/luminosities at two wavebands, i.e., $\alpha_{\rm ox} \equiv -\log(L_{\rm X}/L_{\rm opt})/\log(\nu_{\rm x}/\nu_{\rm opt})$. Astronomers usually take $1.2\times10^{15}$ Hz (wavelength $\lambda\approx 2500$ \AA) and $5\times10^{17}$ Hz (photon energy $h\nu\approx 2$ keV) as representative frequencies of (blue-band) optical and X-rays, respectively. For our adopted $\dot{m}_{\rm bin}$ and simplified model that does not consider any corona above SSD (but see e.g., disk-corona model, \citealt{Haardt1991, Kubota2019}), only when the \texttt{pri-disk} becomes an HAF, will there be notable emission above 2 keV. Thermal emission from either SSD or Slim, has a negligible contribution in hard X-rays.

To summarize, there are several important consequences in the outcome SED due to the existence of a secondary massive BH. First, most existing observations in optical (and infrared for those high-redshift sources) may underestimate the actual mass supply to the central engine(s) by a factor of several, because these observations only measure accretion rate onto the primary BH, but miss that onto the secondary BH. This may also provide one possible explanation for the diverse AGN luminosities in galaxies with similar nuclear gas properties that may provide similar gas supply to two BHs $\dot{M}_{\rm bin}$. Recently, pulsar timing array has reported gravitational wave detections at nHz frequencies, whose origin is likely stochastic inspiralling (i.e., small $a_{\rm sep}$) binary SMBHs (e.g., \citealt{NANOGrav2023, CPTA2023}). Interestingly, \citet{Sato-Polito2023} reported a tension that, the amplitude of the detected gravitational wave signal requires inspiralling SMBHs that are approximately ten times heavier (in mass) than what estimated by Soltan's argument.Our investigation may provide a clue to this discrepancy. If these inspiralling SMBHs typically have a mass ratio of $q\sim 0.02-0.3$ and are fed with sufficient gas (e.g., nearly Eddington supply, $\dot{m}_{\rm bin}\sim 1$, quite reasonable at redshift $z\ga 2$), then the observed radiation can be highly suppressed because the \texttt{sec-disk} enters Slim state (cf. Fig.~\ref{fig:efficiency}). In this case, Soltan's argument with $10\%$ radiative efficiency underestimates the actual BH mass at these redshifts.

Second, even without considering the phenomenological soft X-ray excess component or corona component as in most AGN studies, the UV up to soft X-rays is still highly changed. Clearly the thermal emission from \texttt{sec-disk} whose $q\sim 3\times 10^{-3} - 5\times 10^{-2}$ (the gap-CBD transition case) is somewhat equivalent to a soft X-ray excess, considering that thermal emission peaks at around $h\nu \propto kT_{\rm SSD}\propto\mbh^{-1/4}\dot{m}^{1/4}$ \citep{Frank2002}, and the secondary BH has a smaller $\mbh$ together with a higher $\dot{m}$.

\begin{figure*} 
    \centering
    \includegraphics[width=0.45\textwidth]{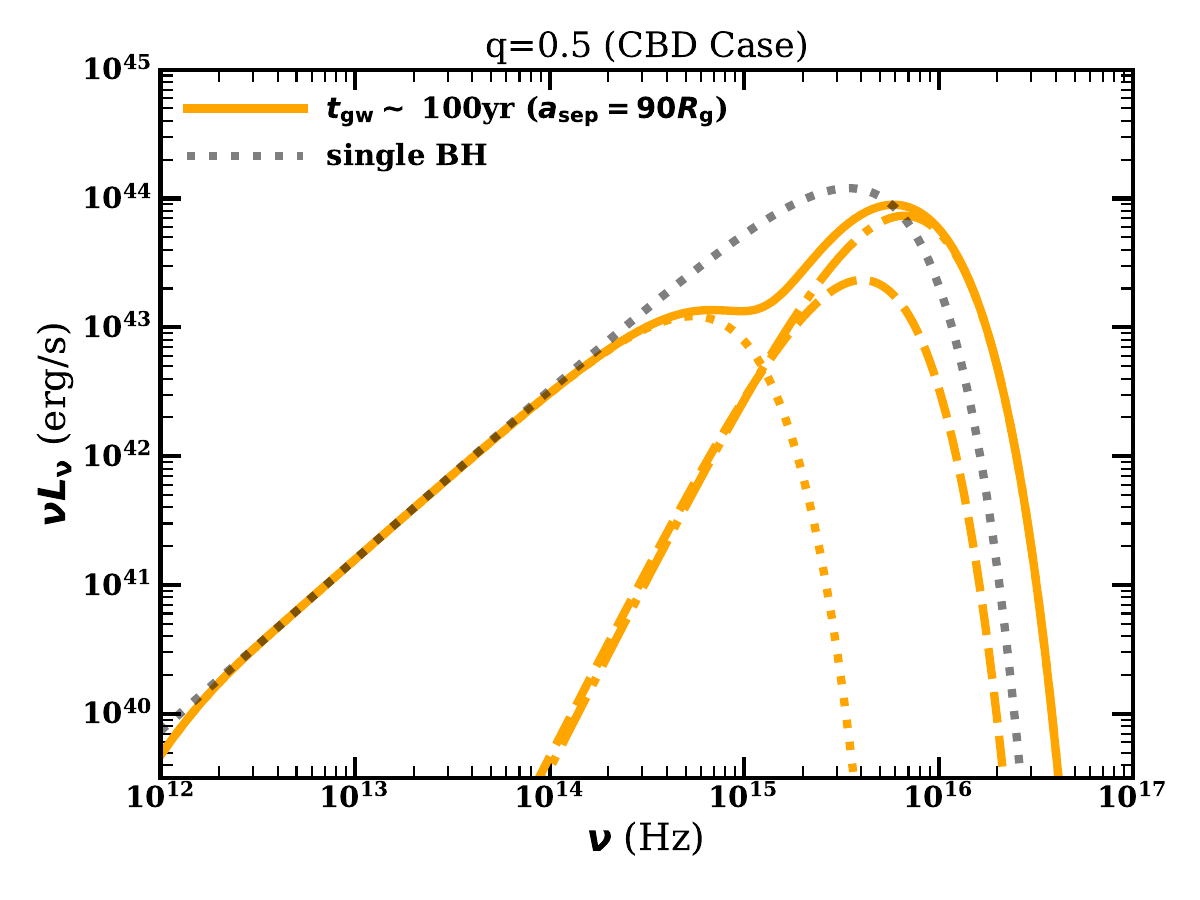}
    \includegraphics[width=0.45\textwidth]{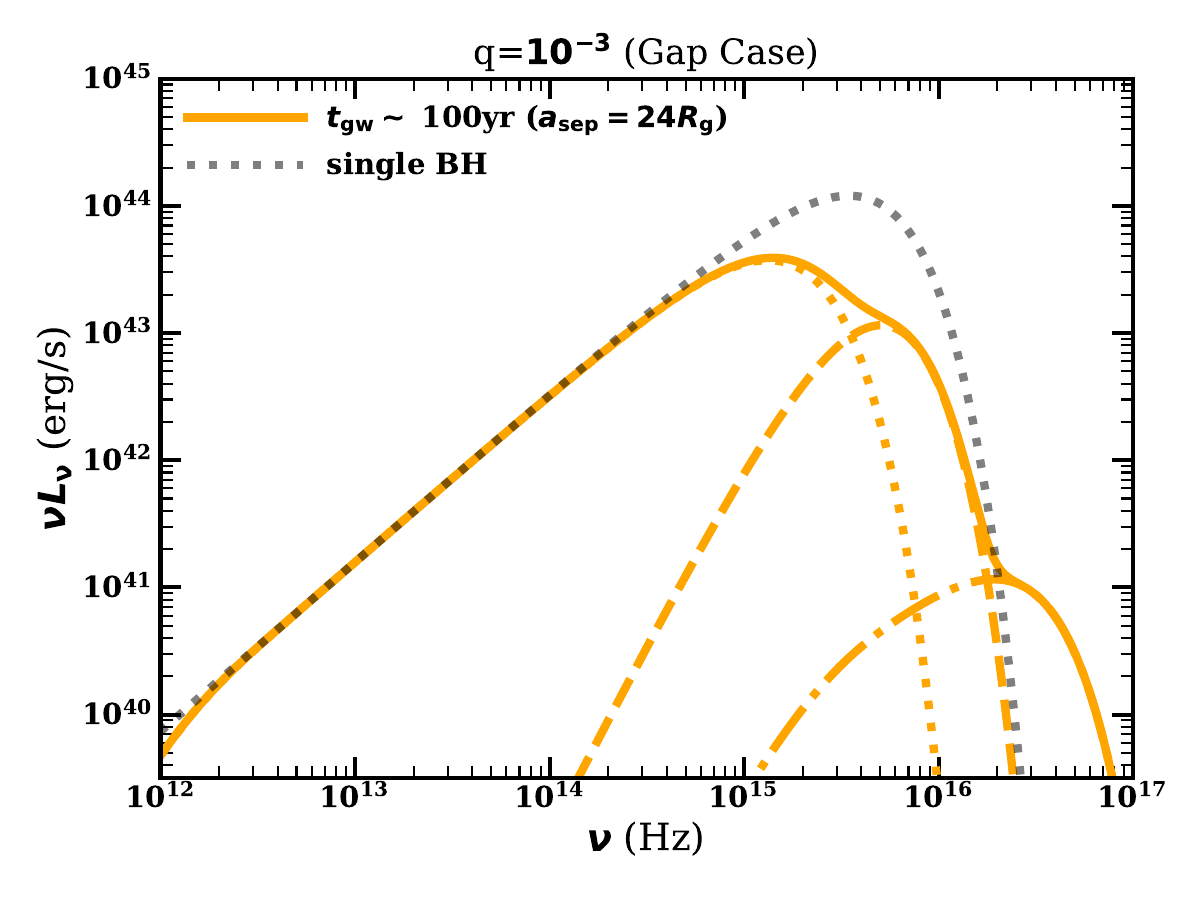}\\
    \includegraphics[width=0.3\textwidth]{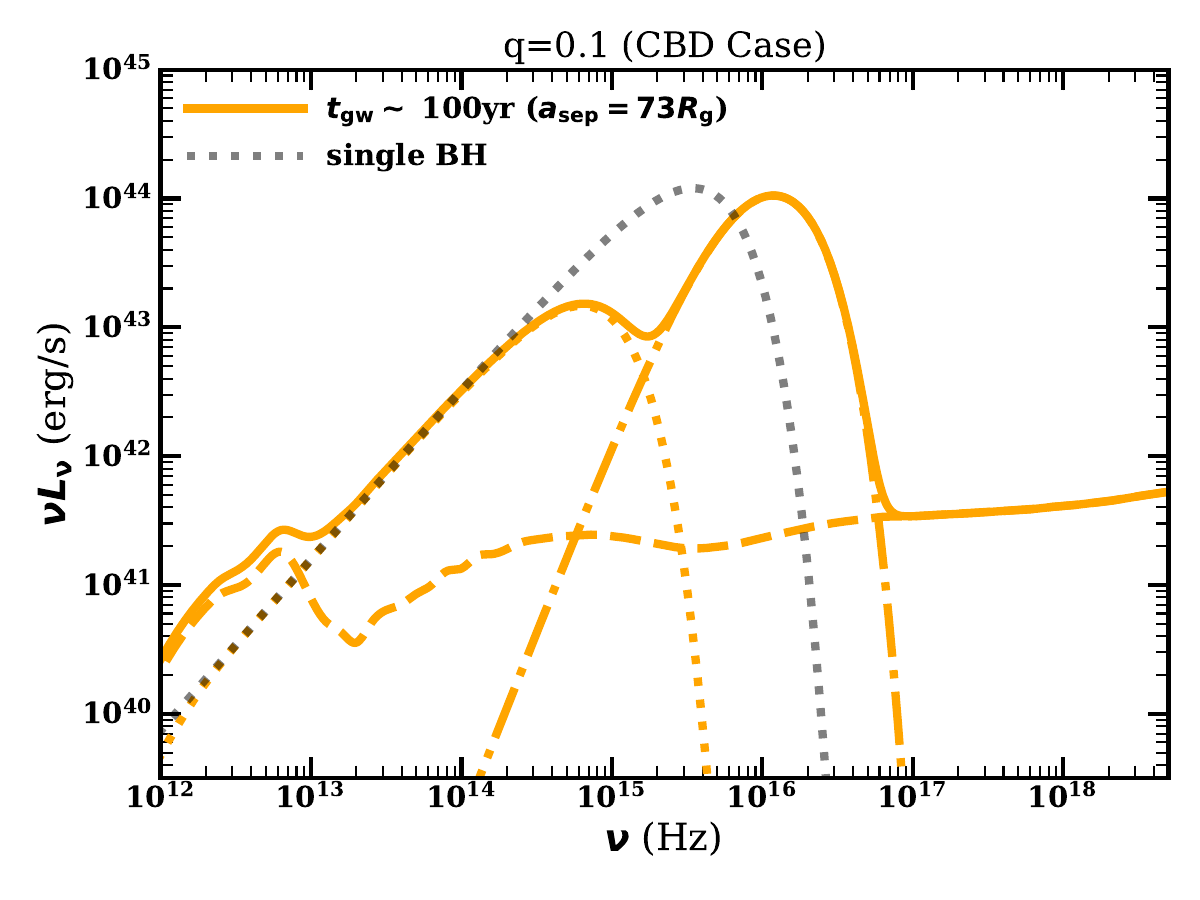}
    \includegraphics[width=0.3\textwidth]{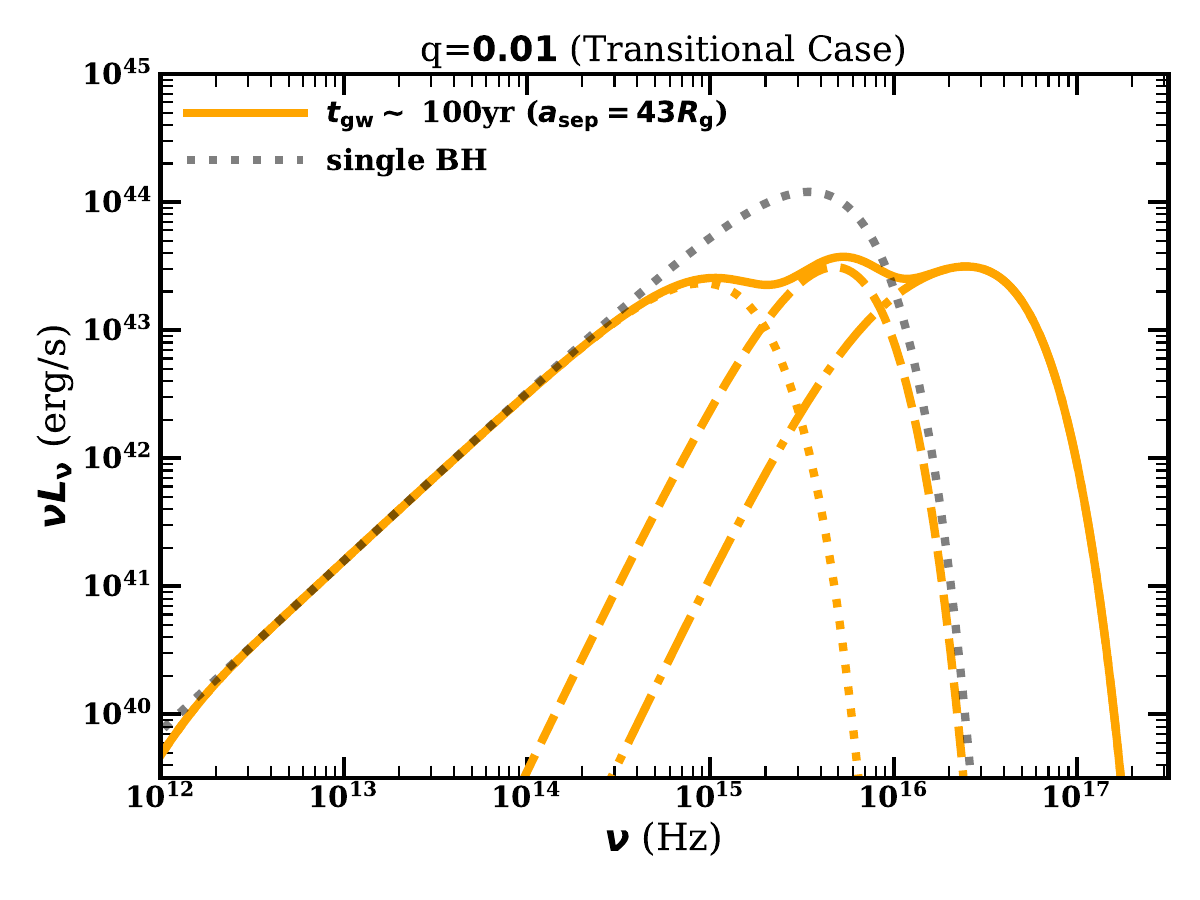}
    \includegraphics[width=0.3\textwidth]{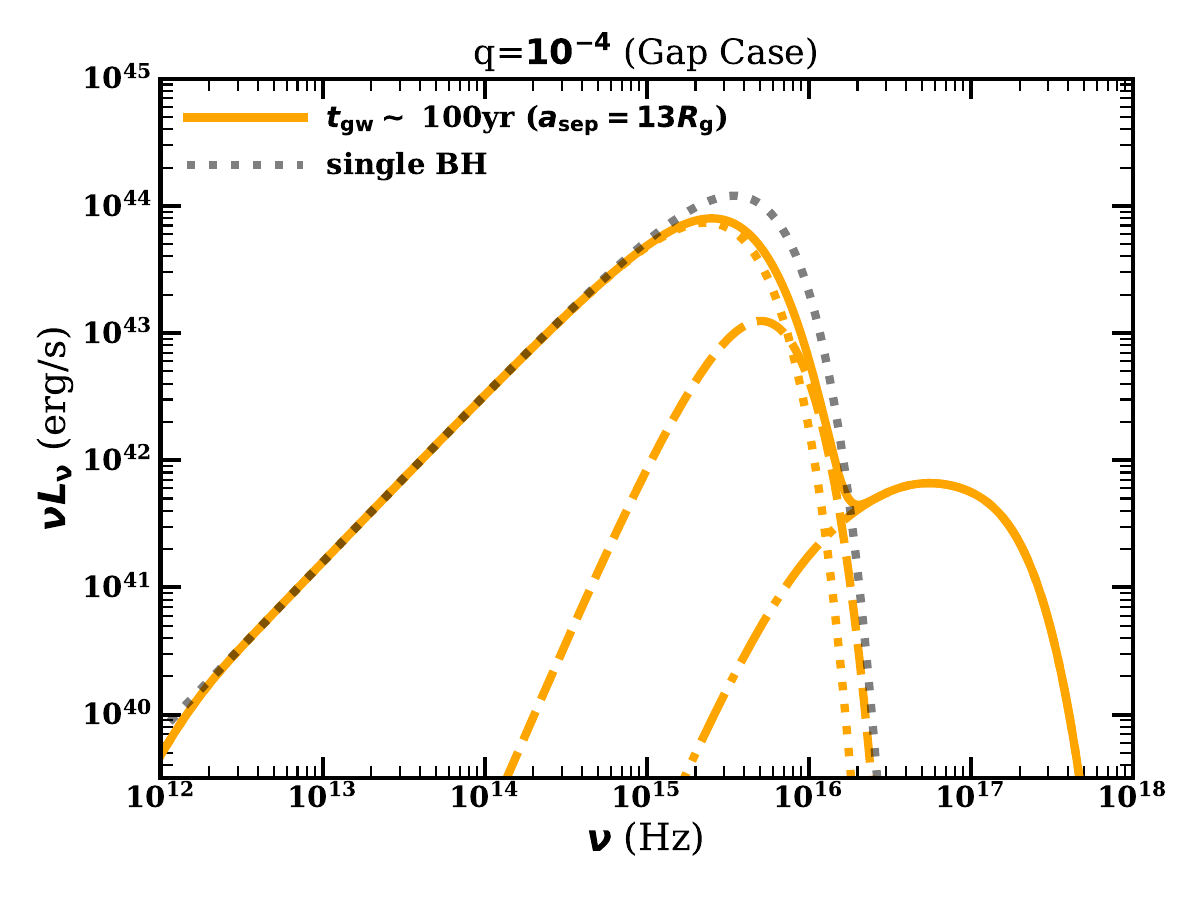}
    \caption{SEDs for binary massive BH systems whose merger time is approximately 100yr. The total BH mass of binary BHs is $M_{\rm bin}=10^8\msun$, and the total accretion rate $\dot{m}_{\rm bin}=0.1$. On top of each panel label the mass ratio of the binary BH system, i.e. $q = 0.5$ (top left, CBD case), $10^{-3}$ (top right, gap case), $0.1$ (bottom left, CBD case), $0.01$ (bottom middle, transitional case), and $10^{-4}$ (bottom right, gap case). The separation of two BHs $a_{\rm sep}$ are labeled on each panel. The line-style conventions used to represent the different disk components and accretion states are consistent with those established in Fig. \ref{fig:q=0.5}.}
    \label{fig:meger_two}
\end{figure*}

\begin{figure*}
\centering
  \includegraphics[width=0.9\textwidth]{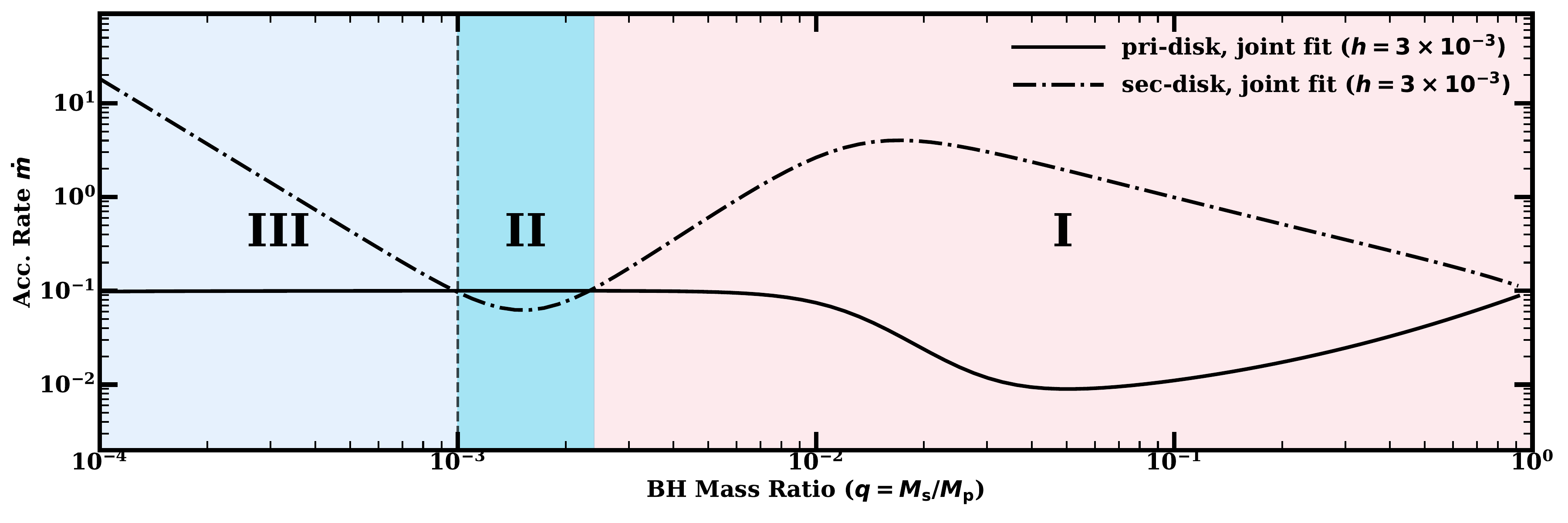}\\
  \includegraphics[width=0.9\textwidth]{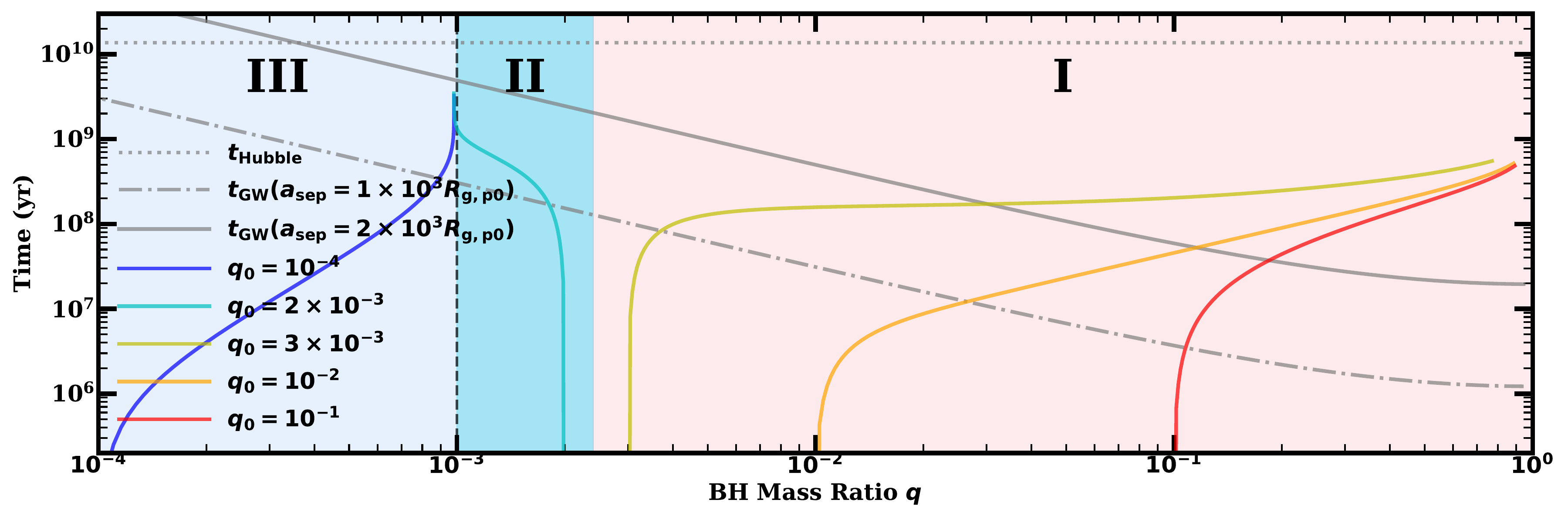}
  \caption{{\it Top panel}: Accretion rates of the primary mini-disk (\texttt{pri-disk}, dot-dashed curve) and the secondary mini-disk (\texttt{sec-disk}, solid) as a function of the binary mass ratio $q$.
  {\it Bottom panel}: Evolutionary tracks of the mass ratio $q$ for different initial values, i.e. $q_0=1\times10^{-4}$ (blue curve), $2\times10^{-3}$ (cyan), $3\times10^{-3}$ (yellow), $0.01$ (orange), $0.1$ (red), respectively. The and dot-dashed and solid gray lines show the gravitational-wave merger timescale as a function of $q$ for binary separations $a_{\rm sep}=1\times10^3 R_{\rm g,p0}$ and $2\times10^3 R_{\rm g,p0}$. In both panels, we fix $\dot{m}_{\rm bin} = 0.1$, and adopt the fitting function of $\lambda(q)$. We define three regimes based on the condition $\dot{m}_{\rm s} > \dot{m}_{\rm p}$. In Regimes I and III, $\dot{m}_{\rm s} > \dot{m}_{\rm p}$, so $q$ increases with time; in Regime II, $\dot{m}_{\rm s} < \dot{m}_{\rm p}$, so $q$ decreases with time. The unique presence of Regime II gives rise to a distinct equilibrium solution where $q\sim10^{-3}$, whereas the more massive binary tends to evolve toward an equal mass ratio. The horizon dotted line in the bottom panel shows the Hubble time $t_{\rm Hubble}=13.7$ Gyr.}
  \label{fig:acc_time}
\end{figure*}

\subsection{SEDs of accretion onto pre-merger binary massive BHs}

As the binary black holes draw closer to each other, the merger of binary BHs is predominantly driven by gravitational wave emission, the merger timescale can be described by the Peters formula \citep{Peters1964}:
\begin{eqnarray}
    t_{\rm gw} & = &\frac{5}{256} \frac{a_{\rm sep}^4c^5}{G^3M_{\rm p} M_{\rm s}(M_{\rm p}+M_{\rm s})}.\nonumber\\
    & = & \frac{5}{256} \frac{(1+q)^2}{q} \frac{R_{\rm g}}{c} \left(\frac{a_{\rm sep}}{R_{\rm g}}\right)^4.\label{eq:merger}
\end{eqnarray}
In this regime, gravitational-wave radiation is significantly enhanced as the dominant driver of the merger process, with the potential for a direct detection of the gravitational wave. At the same time, the detection of electromagnetic counterparts from merging binary massive BHs on short timescales can enhance source localization when combined with gravitational wave observations (e.g., \citealt{Graham2023}).

In this section, we follow Eq.~(\ref{eq:merger}) to adjust the orbital separation $a_{\rm sep}$ such that the two BHs will merger in approximately $100$ years. This typically corresponds to a separation of tens to hundreds of gravitational radii of the binary. Clearly when two BHs become even closer, the \texttt{cb-disk} will detach from the evolution of binary massive BHs. Here we take this near-critical separation, and under the assumption that the accretion disk structure can remain stable during this pre-merger phase, to investigate the spectral properties of binary massive BH systems across different mass ratios.

As analyzed in the following Section ~\ref{sec:evol}, binary massive BHs tend to evolve toward two distinct mass-ratio regimes prior to merger: a low-mass-ratio regime with $q\sim10^{-3}$, and a comparable-mass regime with $q\sim1$. For our chosen $M_{\rm bin}=10^8\msun$, the 100 yr merger timescale corresponds to $a_{\rm sep} = 90 R_{\rm g}$ for the $q=0.5$ CBD case and $a_{\rm sep} = 24 R_{\rm g}$ for the $q=10^{-3}$ gap case. These two situations are illustrated by the top-left and top-right panels of Figure~\ref{fig:meger_two}. In both cases, the gap/cavity lies close to the primary BH, producing a spectral notch in the UV band. A significant difference in extreme-UV ($3\times10^{15}-3\times10^{16}$ Hz, or equivalently, $100-1000$ \AA) is observed, i.e. very hard if $q=0.5$ but extremely soft if $q=10^{-3}$.

We now turn to other cases, which are shown in the bottom panels of Fig.~\ref{fig:meger_two}. For the CBD case of $q = 0.1$ (bottom left panel of Fig.~\ref{fig:meger_two}), we notice that the \texttt{pri-disk} is now a HAF. Because of a smaller $a_{\rm sep}$, the \texttt{pri-disk} is also small in size. Consequently, its total outflow strength is reduced, and the emission from HAF, especially that spanning from the X-rays to gamma-rays, is significantly enhanced. For the transitional case of $q=0.01$ (bottom middle panel of Fig.~\ref{fig:meger_two}), we find that all the three disks share similar bolometric luminosities, but prominent in different wavebands. Consequently, it creates a fairly flat spectrum between $6\times10^{14}$Hz and $4\times10^{16}$Hz (optical-to-UV). For the gap case of small mass ratios $q$ (i.e. $q=10^{-4}$, bottom right panel of Fig.~\ref{fig:meger_two}), the close separation of the two black holes suppresses emission from the \texttt{pri-disk}, while the emission from the \texttt{sec-disk} is in X-rays due to its high accretion rate (in Eddington unit) and small $\mbh$ (see also Fig.~\ref{fig:SED_three}). Clearly, such system has an extremely soft optical-to-X-ray spectrum, with the spectral index $\alpha_{\rm ox} > 3$.

\section{ Evolution of Binary Massive Black Holes}
\label{sec:evol}

We now investigate the evolutionary pathways (over cosmic time) of binary massive BHs, where the accretion rate redistribution among two BHs plays a pivotal role. For this investigation, different initial conditions are considered, in order to probe their future trajectory of the binary system.

The top panel of Figure~\ref{fig:acc_time} is a re-plot of Fig.~\ref{fig:acc_disktypes}. Two intersection points at $\dot{m}_{\rm s} = \dot{m}_{\rm p}$ are identified, one at $q \approx 1\times10^{-3}$ and the other at $q_{\rm crit} \approx 2.5\times10^{-3}$. These two mass ratios are of particular importance for the evolution of binary massive BHs, as they determine the evolutionary pathways of binaries with different initial mass ratios. From the definition of $q$, we have (note that $\dot{M}_{\rm Edd}\propto \mbh$),
\begin{eqnarray}
\frac{dq}{dt} & = & \frac{\dot{M}_{\rm s}}{M_{\rm p}} - \frac{M_{\rm s}\dot{M}_{\rm p}}{M_{\rm p}^2} = \frac{M_{\rm s}}{M_{\rm p}}\left(\frac{\dot{M}_{\rm s}}{M_{\rm s}}-\frac{\dot{M}_{\rm p}}{M_{\rm p}}\right) \nonumber\\
& = & \frac{q\,(\dot{m}_{\rm s} - \dot{m}_{\rm p})}{\tau_{\rm Salpeter}},\label{q_evol}
\end{eqnarray}
where $\tau_{\rm Salpeter} \equiv M_{\rm BH}/\dot{M}_{\rm Edd} \approx 4.4\times10^{7}$ yr is the Salpeter e-folding BH mass growth timescale \citep{Salpeter1964}. Clearly, the mass ratio $q$ increases only when $\dot{m}_{\rm s} > \dot{m}_{\rm p}$. Based on the above two intersection points, we can divide the space into three regions, which are marked as regions I/II/III in this plot. For region II where $1\times10^{-3} < q < 2.5\times10^{-3}$, the secondary grows slower than the primary ($\dot{m}_{\rm s} < \dot{m}_{\rm p}$), thus we expect the mass ratio $q$ to evolve toward smaller values. In contrast, for region I and III, the mass ratio increases with time. With a fixed value of $\dot{m}_{\rm s} - \dot{m}_{\rm p}$, an analytical expression of the exponential evolution of $q$ can be derived, i.e.,
\begin{equation}
    q(t) = \exp\left({\frac{\dot{m}_{\rm s} - \dot{m}_{\rm p}}{\tau_{\rm Salpeter}}t}\right). \label{eq:q_evol2}
\end{equation}
From this formulae, we clearly find that the actual timescale of $q$ is proportional to $|\dot{m}_{\rm s} - \dot{m}_{\rm p}|^{-1}$.

As shown in the bottom panel of Figure~\ref{fig:acc_time}, we now compute the cosmic mass growth history of each black hole for systems with initial ratio $q_0 = 10^{-4}$, $2\times10^{-3}$, $3\times10^{-3}$, $0.01$, and $0.1$. Here we adopt an initial primary BH mass $M_{\rm p,0} = 10^8\msun$ and a constant (time-independent) binary accretion rate $\dot{m}_{\rm bin}(t) \equiv 0.1$ (note that the actual accretion rate $\dot{M}_{\rm bin}$ still varies with increasing $M_{\rm bin}$). For systems in both regions III and II, the long-term binary evolution converges toward $q \sim 1\times10^{-3}$. Conversely, for systems in region I, with an initial mass ratio $q_0 > q_{\rm crit}$, the binary evolves toward mass equality ($q \sim 1$). We caution that the exact time of these evolutions has a negative dependence on $\dot{m}_{\rm bin}$.

The bottom panel of Fig.~\ref{fig:acc_time} shows the evolutionary tracks of the mass ratio $q$ for different initial values, i.e. $q_0=1\times10^{-4}$ (blue curve), $2\times10^{-3}$ (cyan), $3\times10^{-3}$ (yellow), $0.01$ (orange), $0.1$ (red), respectively. As learned from Eq.~(\ref{eq:q_evol2}), $q$ evolves most rapidly at a large value of $|\dot{m}_{\rm s} - \dot{m}_{\rm p}|$, i.e. $q \lesssim 4\times10^{-4}$ or $ 6\times10^{-3} \lesssim q\lesssim 0.15$. This is also clearly demonstrated in this plot. For a system with an initial mass ratio $q_0=3\times10^{-3}$, it takes about $1.4\times10^8$ yrs to reach $q=6\times10^{-3}$ (increases by only a factor of 2). On the other hand, it takes about $1\times10^8$ yrs to evolve from $q=6\times10^{-3}$ to $q=0.2$ (increases by a factor of $\sim 30$).

For comparative purpose, we also show in the bottom panel of Fig.~\ref{fig:acc_time} the gravitational wave-driven inspiral timescale as a function of mass ratio $q$, where the binary separations are $a_{\rm sep}=2000R_{\rm g,p0}$ (gray solid) and $1000 R_{\rm g,p0}$ (gray dot-dashed), respectively. Here $R_{\rm g,p0}$ is the gravitational radius of the primary black hole. $t_{\rm GW}$ may exceed the Hubble time of the Universe $t_{\rm Hubble}=13.7$ Gyrs at small-$q$ end.

We also compute the gravitational migration timescale of binary systems (not shown here due to uncertainties in its estimation). We first focus on the gap case. According to Eq.~(29) in \citet{Kanagawa2018}, the migration timescale in the gap case is significantly shorter than the mass-growth timescale, with $t_{\rm mig}\sim10^7$yr for a mass ratio of $q=10^{-3}$. However, the migration timescale in AGN disks may be considerably longer and is subject to substantial uncertainties, depending on the disk structure \citep{Li2025}. These uncertainties arise from  over-simplified assumptions in existing models (e.g., \citealt{Kanagawa2018, Lai2023}). If we temporarily set aside these uncertainties in the migration timescale, a clearer picture emerges under the condition that for $q\lesssim10^{-2}$, the merger timescale $t_{\rm gw}$ is much longer than the mass-growth timescale. In this regime, accretion can efficiently drive significant evolution of the mass ratio until the binary merges. Consequently, for initial mass ratios $q \lesssim \text{a few} \times 10^{-3}$, the mass ratio evolves toward an equilibrium value $q \sim 10^{-3}$.

In contrast, for the CBD case, the migration is considerably slower, with $t_{\rm mig}\sim 4\times10^8$ yr, exceeding the mass-growth timescale (see Eq.~17 in \citealt{Lai2023}) and merger timescale for $a_{\rm sep}=2000R_{\rm g,p0}$ (see Fig.~\ref{fig:acc_time}). These findings suggest that the majority of binary SMBHs with mass ratios $q\lesssim10^{-2}$ are likely to merge before they can evolve into systems with $q=10^{-3}$ or large mass ratios $q\sim1$. 

\section{Discussions}\label{sec:discussion}

In this work, we focused on two types of ``interacting'' binary massive BH systems, one is SMBH-IMBH accretion system, and the other is SMBH-SMBH system. Moreover, the prerequisite of this investigation is that, the two massive BHs need to be ``interacting'', thus the separation should be less than $0.1$ pc. As addressed in {\it Introduction}, there are various mechanisms to form such systems at the center of galaxies. The mass ratios in this work reflects this late-time evolved $q$ values in this stage, thus are quite uncertain now. For example, it remains inclusive even for the existence of IMBH in dense global clusters. For the more recognized galaxy-galaxy merger pathway, various post-process methods are implemented to understand the (un-resolved) evolution of massive BHs based on modern cosmological simulations (e.g., \citealt{Volonteri2020}). Some works suggest that more fraction of binary SMBHs are in the range $q\sim 0.1-1$ when they are in ``interacting'' binary accretion state. However, these investigations adopt the conventional picture of mass accretion rate redistribution in the gap case, which should be treated with caution.

\subsection{Observational Implications}

\begin{table}
  \begin{threeparttable}
  \caption{Doppler Effect by Orbital Motions of BH ($q\lesssim 0.05$ Case)} \label{tab:doppler}
  \begin{tabular}{ccccc} 
  \hline
  $\asep(R_{\rm g})$ & $P_{\rm bin}$ (yr) & $V_{\rm sec}$ (cm/s) & $\Delta F_{\nu}/F_{\nu}$ & $\Delta\nu/\nu$ \\
  \hline
  50 & 0.035 & $4.2\times10^9$ & 0.21 & 0.065 \\
  $10^2$ & 0.98 & $3.0\times10^9$ & 0.15 &0.047 \\
  $2\times10^3$ & 8.8 & $6.7\times10^8$ & 0.034 & 0.011 \\
  $10^4$ & 97 & $3.0\times10^8$ & 0.015 & 0.0050\\
  \hline
  \end{tabular}
  \begin{tablenotes}
  \small
  \item 1. Total BH mass is fixed to $M_{\rm bin} = 10^8 \msun$.
  \item 2. For larger BH mass ratios, the orbital velocity of the primary and the secondary BHs at the same $a_{\rm sep}/R_{\rm g}$ are reduced by a factor of, respectively, $\sqrt{q/(1+q)}$ and $1/\sqrt{1+q}$, compared to $V_{\rm sec}$ reported here. $\Delta F_\nu/F_\nu$ and $\Delta \nu/\nu$ can also be re-calculated accordingly.
  \end{tablenotes}
  \end{threeparttable}
\end{table}

\subsubsection{Doppler Effect}\label{Doppler}


Long-term monitoring of binary SMBHs enables the detection of periodic brightening in the observed flux, a signature of the orbital motion of the secondary BH. Recent studies have reported periodic blazars in the optical/near-infrared (e.g., \citealt{Graham2015}) and gamma-ray (e.g., \citealt{Ackermann2015, Sandrinelli2016}) bands. However, periodic flux variations can arise from multiple mechanisms. For example, in periodically loud radio quasars, models such as jet precession can also produce periodic variability \citep{Begelman1980}. Therefore, it is essential to distinguish whether the observed periodic flux modulation is caused by binary SMBHs. Here, we estimate the modulatory effects of the Doppler effect on frequency and flux intensity for binary massive BHs with $M_{\rm bin}=10^8\msun$.

Relativistic Doppler boosting factor is (e.g., \citealt{Charisi2018}),
\begin{equation}
    \Delta F_{\rm \nu}/F_{\rm \nu}=\pm \left( 3-\alpha \right) \left(\beta\cos \phi \right) \sin i,
\end{equation}
where $i$ and $\phi$ are the orbital inclination and orbital phase, respectively. $\alpha$ is the spectral index (here defined as $F_\nu\propto\nu^{\alpha}$) which for the emission in optical band is typically in the range of $0-3$. $\beta$ is the orbital velocity in unit of c. The dynamical motions of primary/secondary BH will lead to periodic variations in observed flux. Considering the difference in respective SEDs among \texttt{pri-disk} and \texttt{sec-disk} (see Sec.~\ref{sec:rad}), a careful selection of observational waveband is crucially important. We first consider $q \lesssim 0.05$. In this case, the primary SMBH is nearly static and thus introduces weak periodic variation in $\Delta F_{\rm \nu}/F_{\rm \nu}$. The secondary BH, on the other hand, has $V_{\rm sec} = \Omega_{\rm bbh} R_{\rm sec} \approx \left(\frac{GM_{\rm bin}}{a_{\rm sep}}\right)^{1/2}$, and its emission is more evident at the extreme UV band ($\ga 10^{16}$ Hz. It may move to near-UV band for systems at a high-redshift). The Doppler amplification factor is summarized in Table \ref{tab:doppler}. The maximum boost factor $\Delta F_{\nu}/F_{\nu}$ is generally less than $20\%$. When the observed periodic flux enhancement falls within this range, it may be attributed to Doppler boosting from the orbital motion of a binary SMBH system \citep{Zheng2016}. For systems with comparable BH mass (i.e. $q\sim 0.3-1$) and comparable luminosities, we should observe two groups of periodic variations with the same period ($P_{\rm bin}$), but differ exactly by a phase of $\pi$ (orbital phase of the primary BH differs to that of the secondary by $|\phi_{\rm pri} - \phi_{\rm sec}|=\pi$). In this case, the overall variation is dominated by \texttt{sec-disk} emission (cf. Fig.~\ref{fig:lbol_q}), and disentangling the relatively weaker \texttt{pri-disk} emission will require more detailed modeling efforts.

AGNs generally have a broad-emission-line region located at $\sim 10^2 - 10^{4} R_{\rm g}$ \citep{Padovani2017}, among which some may host two massive BHs at their nuclei region. The dynamics of broad-line-region is highly unclear for binary SMBH systems whose $q\sim 0.3-1$. For those binary massive BH systems whose $q\lesssim 0.05$, we may aggressively assume that part of the broad-line region gases very close to the secondary BH may be dragged and co-moves with the secondary BH. In this case, we may observe a periodic frequency shift of individual emission lines by these broad-line region gases. The frequency shift can be expressed as,
$\Delta \nu/\nu=\sqrt{ \left( 1-\beta^2 \right)}/\left ( 1-\beta\cos i\right )-1$. The maximum frequency shift in these $q\lesssim 0.05$ system is provided in Table \ref{tab:doppler}, which is generally less than $7\%$. One notable advantage compared to continuum flux variation above is that, these broad lines are measured through spectroscopic observations in the more common optical band. Again, we caution that, due to the weak potential of secondary BH, only a limited portion of gases of the broad-line region can be affected by the secondary BH. The strength of these periodically shifted emission lines should be fairly weak (especially when $q\lesssim 0.01$), thus highly suspicious for a reliable detection in a foreseeable future.

\subsubsection{Observational Signatures}

Our results indicate that, in the CBD case with large mass ratios, the \texttt{sec-disk} produces emission in UV and soft X-rays. However, due to strong absorption in these wave bands, such excess is difficult to observe by current facilities (but see the potential capabilities of UVEX that covers $\sim 1400-2700$ \AA, cf. \citealt{Kulkarni2021}). For small mass ratios ($q\sim10^{-4}$), the \texttt{sec-disk} attains a higher accretion rate and emits in the hard X‑ray band (see Fig.~\ref{fig:SED_three}), make it more  detectable. Notably, at such mass ratios the \texttt{sec-disk} enters a super‑Eddington Slim state. Its X-ray luminosity is highly suppressed, yielding a flux roughly two orders of magnitude below that of that of the \texttt{pri-disk}. Besides, hot corona above \texttt{pri-disk} (SSD), which is totally neglected in this work, may contaminates the hard X-rays.

For mass ratios $q\gtrsim10^{-3}$, a significant spectral notch feature in the NIR-UV bands is predicted (e.g., \citealt{Roedig2014}). The exact location and the width of the notch depends on the separation of two BHs. For nearby Seyferts the UV band is severely absorbed; whereas for high-redshift ($z\ga 3$) AGNs with small $a_{\rm sep}$, this wavelength range is redshifted to the optical band thus is easier to observe. Based on our analysis in Sec~\ref{sec:evol}, a majority of binary massive BHs with mass ratios $q \lesssim 10^{-2}$ are likely to merge before they can evolve into large-mass-ratio systems. This finding may explain why we do not have much strong observational evidence on the spectral notch feature in Seyferts nor in high-redshift AGNs (but see below for a discussion on ``Little Red Dots'' as observed by JWST).

Several candidates of binary SMBHs with notch features are reported in literature. For example, \citet{Yan2015} proposed that the observed flux deficit at $\lambda\sim 4000-2500$ \AA of the optical–UV spectrum of Mrk 231 is likely due to the existence of a secondary black hole. Conversely, \citet{Farris2015} derived the thermal spectrum of a binary SMBH system directly from numerical simulations. Their findings indicate that hot gas within mini-disks and accretion streams inside the cavity generates a spectral surplus—rather than a deficit—in the high-energy regime.

Recently JWST discoveries an un-expected population of ``Little Red Dots'' (LRDs, \citealt{Labbe2024, Labbe2025}, and references therein) at $z>4$ with a number density comparable to luminous quasars. These sources ubiquitously exhibit V-shaped spectra that morphologically resemble the notch features predicted for accreting binary SMBHs. However, their physical origins are fundamentally different. In LRDs, the notch is anchored at the hydrogen Balmer break (e.g., \citealt{Labbe2024, Baggen2024}) at a fixed rest-frame wavelength of $\sim 3600\,\text{\AA}$. In contrast, the binary-induced notch is determined by the orbital separation and lacks a fixed rest-frame frequency, although it can coincidentally fall at the same frequency for specific orbital configurations. Nevertheless, the accreting binary SMBHs considered here may act as contaminants — and thus byproducts — in systematic photometric or spectroscopic searches for LRDs \citep{Inayoshi2026}.

Beyond Doppler boosting, the binary’s gravitational potential periodically perturbs the accretion flow, generating a density lump close to the edge of the cavity (e.g., \citealt{D'Orazio2015}), modulating the local dissipation rate and temperature distribution. This mechanism can produce variability amplitudes of order $\sim$10-100 and exhibit quasi-periodic behavior linked to $\sim5 P_{\rm bin}$ (see e.g., Figs.~3 and 4 in \citealt{Lai2023}). \citet{Tiede2025} compares the cooling timescale of various radiative processes with the binary orbital period. They find that thermal emission from SSD and low-frequency synchrotron emission from HAF show strongest periodic modulation. The former benefits from short thermal timescales, while the latter arises from rapid radiative cooling. In contrast, other thermal mechanisms have significantly longer cooling times, which smear and suppress the periodic signal. These timing characteristics help distinguish modulation caused by binary orbital motion from other periodic AGN phenomena, such as jet precession or quasi-periodic oscillations (e.g., \citealt{Gierlinski2008, Britzen2018}).

Our results also show that, for the given $\dot{m}_{\rm bin} = 0.1$, the \texttt{pri-disk} becomes an HAF at $q\approx0.1$. Compton up-scattering in the HAF produces high-energy photons in the X-ray and $\gamma$-ray bands. However, in lower-luminosity AGNs, a truncated-disk configuration--e.g., an outer SSD plus an inner HAF--can produce a similar broadband SED, requiring additional multi-wavelength or variability constraints to confirm a binary origin (e.g., \citealt{Dewangan2021, Stolc2023}).

In addition, although our current study focuses on binary massive BHs with a typical total mass of $10^{8} \msun$, the theoretical framework presented here can be naturally extended to even more massive systems, such as the $\sim 6\times10^9 \msun$ SMBH in the giant elliptical galaxy M87. As discussed by \cite{Davari-etal.2017}, giant ellipticals should have experienced at least one major merger (i.e., $q \ge 0.25$) since $z = 2$, during their structural transformation from compact red nuggets into local giant ellipticals. In the context of our accretion models, a major merger with $q \ge 0.25$ falls squarely into the CBD regime. During this active phase, the binary system is expected to exhibit the characteristic spectral "notch" across the infrared to ultraviolet range, alongside a prominent radiative contribution from the secondary mini-disk. Scaling our SED predictions to the $10^9 \msun$ mass regime would provide valuable electromagnetic diagnostics for identifying major SMBH mergers in the high-redshift universe, ultimately shedding light on the co-evolution, accretion, and assembly history of giant ellipticals like M87.

Finally, simulations of circumbinary disk indicate that the merger event can produce observable electromagnetic emission before, during, and after coalescence \citep{Graham2020, Kocsis2011, Haiman2008, Haiman2009, Farris2015b, Haiman2017}. This will crucially assist in pinpointing the gravitational-wave source.

\subsection{Limitations}

In this work, we do not explore the impact of BH spin, but fix the spin of two BHs to a median value of $a^{\star}=0.5$. Higher spin BH has a smaller BH horizon, consequently accretion onto higher spin BHs will produce more radiation. Besides, the peak frequency of SSD emission also shifts toward higher energies. Additionally, through the coupling to the orbital angular momentum, BH spin orientation also modulates the orbital precession and gravitational recoil, thus revises the merger timescale and gravitational wave signals (e.g., \citealt{Apostolatos1994, Campanelli2007}).

In our accretion model, the prescription of accretion rate redistribution among two BHs is taken directly from hydrodynamic simulations of the gap and CBD cases (e.g., \citealt{Fung2014, Kanagawa2017, Duffell2020}). These simulations usually assume a SSD with an aspect ratio of $h\gtrsim0.03$. For binary massive BH systems containing SSD, Slim and HAF, and/or much more thinner SSD with $h\sim 10^{-3}$, further simulations are needed to validate the prescription.

Besides, there are currently no dedicated simulations that concern the mass ratio in transitional state between the gap case and the CBD case. In this work, we adopt a rough criterion that $q_{\rm max,gap}$
which has a gap width of $W_{\rm gap}=1.5a_{\rm sep}$. Uncertainty in $q_{\rm max,gap}$ has a notable impact on the accretion rate estimates for mass ratios in the range $10^{-3}<q\lesssim10^{-2}$. The maximum gap width reached in previous numerical simulations is $\approx 0.8a_{\rm sep}$ \citep{Duffell2020, Kanagawa2017}, whereas we adopt a much broader value, following the analytical gap-width formula Eq.~(\ref{eq:gapwidth}). We notice that the surface density profile inside the gap remains reliable over a broader parameter range. Based on the accretion rate of the gap case \citep{Li2023}, we suggest that the applicable upper limit of their accretion-rate expression likely exceeds $q_{\rm max, gap}$. Indeed, \citet{Fung2014} carried out a series of simulations, among which a model with $\alpha_{\rm vis}=10^{-3}, h=0.05, q=0.01$ does exceed the $q_{\rm max, gap}$ defined by our criterion.

In this work, we only consider simple accretion configurations/models at different accretion rate. It is well known that the hard X-rays can also arise from Compton scattering in a hot corona above the SSD or Slim (\citealt{Kubota2019} and references therein). Such configuration is typically adopted in bright AGNs (and black hole binaries in soft state). Considering corona is beyond the scope of our work. Besides, outflows in HAF is considered. However, its strength is not well-constrained, either observationally nor theoretically. The X-ray and gamma-ray emission would be more brighter than our current prediction if a weaker outflow strength is adopted (e.g., \citealt{Tiede2025}).

\section{Conclusions}\label{sec:conclusion}


In the era of gravitational-wave astronomy, explore the theoretically-expected existence of close binary massive BHs is one of the frontiers of astrophysics. In this work, we systematically investigate the accretion processes and radiative properties of binary massive BH systems across a broad range of mass ratios from $q=0.5$ down to $q=10^{-4}$. Updates in accretion theory (three accretion models, i.e., SSD, HAF and Slim) and binary accretion (differences among gap case and CBD case, redistribution of $\dot{M}$ between two BHs) are employed. Our study focuses on a typical system with a total BH mass of $10^8 \msun$ and an accretion rate of $\dot{m}_{\rm bin}=\dot{M}_{\rm bin}/\dot{M}_{\rm Edd,bin}=0.1$. Our main conclusions are summarized as follows:

\begin{itemize}
\item We construct a framework for accretion onto binary massive BHs, distinguishing between two accretion configurations: the gap case and the CBD case. Distinct features in SED are observed, which vary systematically with mass ratio $q$. Consistent with previous studies, we find that the SEDs display a characteristic ``notch'' --- a broad spectra depression-across the infrared to ultraviolet range for most mass ratios. For systems with $q = 0.1$, the infrared–ultraviolet continuum is dominated by emission from both the \texttt{sec-disk} and the \texttt{cb-disk}. The \texttt{pri-disk} transitions to a hot accretion flow, contributing (soft and hard) X-ray emission. In systems with $q = 10^{-4}$, the \texttt{sec-disk} becomes Slim, producing excess radiation in soft X-rays.

\item Systems with merger timescales as short as approximately 100 years (corresponding to orbital separations $a_{\rm sep} \lesssim 100 R_{\rm g}$) exhibit distinctive spectral signatures. In particular, the \texttt{cb-disk} emission peaks in the optical-to-UV range. The presence of a gap or cavity introduces a characteristic spectral ``notch'' in the far‑UV range, providing a promising diagnostic for identifying binary massive BHs on the verge of merger.

\item By developing a composite fit for the accretion-rate ratio $\lambda(q)$ across the full mass-ratio range, we identify a critical mass ratio of $q_{\rm crit}\sim10^{-3}$. Systems with initial mass ratios  $q\lesssim2.5\times10^{-3}$ will evolve toward this critical value $q_{\rm crit}$, whereas those above it tend to evolve toward mass equality ($q\rightarrow1$). This bifurcation in evolutionary pathways has important implications for the long-term dynamics of binary massive BH systems in galactic nuclei. It is worth noting that, if we ignore the migration timescale, most binay BHs with mass ratio below $\text{a few}\times 10^{-3}$ may finally merge with a critical mass ratio of $q\sim10^{-3}$.

\item We also predict detectable periodic modulations in binary BH accretion light curves, especially the Doppler‑boosting signals with amplitudes up to $20\%$. Such timing signature, together with the notch‑like SED features, provide complementary observational avenues for confirming binary massive BHs candidates.

\end{itemize}

This study establishes a theoretical framework for interpreting multi‑wavelength continuum observations of AGNs and identifying candidate accreting binary massive BH systems. The predicted spectral features --- particularly the infrared-to-UV ``notch'' and periodic modulation signals—can be tested with existing and forthcoming observational facilities. However, our model relies on several simplified assumptions regarding the disk geometry, and does not consider the highly-magnetized accretion mode (e.g., \citealt{Narayan2003, xie2019}). Future studies should incorporate more sophisticated hydrodynamical simulations to validate these predictions and to investigate the impact of three‑dimensional disk structure and magnetic fields on the accretion flow.

The next generation of space‑based UV telescopes, including proposed missions such as UVEX, LUVOIR and HabEx \citep{Kulkarni2021, LUVOIR2019, Gaudi2018}, and wide‑field time‑domain survey telescopes (e.g., Rubin LSST, \citealt{Brandt2018}; CSST, \citealt{CSST2026}), will provide unprecedented opportunities to test these predictions. These facilities will enable the assembly of large, homogeneous AGN samples monitored with high cadence and broad multi‑wavelength coverage, thereby facilitating systematic searches for the characteristic binary massive BHs signatures predicted in this work. Such observations will not only test the theoretical framework presented here but also yield crucial insights into the population statistics, evolutionary pathways, and cosmic merger rates of accreting binary massive BH systems.

\section*{Acknowledgements}
We appreciate the referee for helpful suggestions. This work was supported in part by the National Natural Science Foundation of China (NSFC Nos. 12373017, 12373070, 12192220 and 12192223), and the Natural Science Foundation of Shanghai (grant No. 23ZR1473700). Y.W. acknowledges the EACOA Fellowship awarded by the East Asia Core Observatories Association. L.C.H. was supported by the National Science Foundation of China (12233001) and the China Manned Space Program (CMS-CSST-2025-A09)



\bibliography{bSMBH-sed.bib}{} 
\bibliographystyle{aasjournalv7}



\end{CJK*}
\end{document}